\documentclass[acmsmall]{acmart}

\usepackage[acronym,toc]{glossaries}

\usepackage{pifont} % Symbols for tables
\usepackage{multirow}
\usepackage{array} % Table sizes
\usepackage{longtable}
\usepackage{booktabs} % pretty tables

\usepackage{rotating}
\usepackage{tabularx}

\AtBeginDocument{%
  \providecommand\BibTeX{{%
    \normalfont B\kern-0.5em{\scshape i\kern-0.25em b}\kern-0.8em\TeX}}}

\setcopyright{acmcopyright}
\acmJournal{CSUR}
\acmYear{2020} \acmVolume{1} \acmNumber{1} \acmArticle{1} \acmMonth{1} \acmPrice{15.00}\acmDOI{10.1145/3427376}

\acmJournal{JACM}
\acmVolume{0}
\acmNumber{0}
\acmArticle{0}
\acmMonth{0}

\usepackage{etoolbox}
\usepackage{soul} 

\newtoggle{finalPaper}

\setstcolor{red}

\toggletrue{finalPaper} % final version

\iftoggle{finalPaper} {

	\newcommand{\rmvtxt}[1]{}}
{

	\newcommand{\rmvtxt}[1]{\st{#1}}}
\newtoggle{removeItalic}
\togglefalse{removeItalic} % regular command
\iftoggle{removeItalic} {
    \renewcommand{\textit}[1]{#1}
}

\begin{document}

%%
%% The "title" command has an optional parameter,
%% allowing the author to define a "short title" to be used in page headers.
\title[Security in BCI: State-Of-The-Art, Opportunities, and Future Challenges]{Security in Brain-Computer Interfaces: State-Of-The-Art, Opportunities, and Future Challenges}

\author{Sergio L\'opez Bernal}
\email{slopez@um.es}
\affiliation{%
  \institution{University of Murcia - Departamento de Ingenier\'ia de la Informaci\'on y las Comunicaciones}
  \city{Murcia}
  \state{Spain}
}

\author{Alberto Huertas Celdr\'an}
\email{ahuertas@tssg.org}
\affiliation{%
  \institution{Waterford Institute of Technology - Telecommunication Software and Systems Group,}
  \city{Waterford}
  \state{Ireland}
}
\affiliation{%
  \institution{University of Zurich - Department of Informatics}
  \city{Zurich}
  \state{Switzerland}
}

\author{Gregorio Mart\'inez P\'erez}
\email{gregorio@um.es}
\affiliation{%
  \institution{University of Murcia - Departamento de Ingenier\'ia de la Informaci\'on y las Comunicaciones}
  \city{Murcia}
  \state{Spain}
}

\author{Michael Taynnan Barros}
\email{michael.barros@tuni.fi}
\affiliation{%
  \institution{University of Essex - School of Computer Science and Electronic Engineering,}
  \city{Essex}
  \state{UK}
}
\affiliation{%
  \institution{Tampere University - CBIG/BioMediTech in the Faculty of Medicine and Health Technology}
  \city{Tampere}
  \state{Finland}
}

\author{Sasitharan Balasubramaniam}
\email{sasib@tssg.org}
\affiliation{%
  \institution{Waterford Institute of Technology - Telecommunication Software and Systems Group,}
  \city{Waterford}
  \state{Ireland}
}
\affiliation{%
  \institution{RCSI University of Medicine and Health Sciences - FutureNeuro, the SFI Research Centre for Chronic and Rare Neurological Diseases}
  \city{Dublin}
  \state{Ireland}
}

%%
%% By default, the full list of authors will be used in the page
%% headers. Often, this list is too long, and will overlap
%% other information printed in the page headers. This command allows
%% the author to define a more concise list
%% of authors' names for this purpose.
\renewcommand{\shortauthors}{L\'opez Bernal, et al.}

% Include the file with the acronyms
\newacronym{BCI}{BCI}{Brain-Computer Interface}
\newacronym{BMI}{BMI}{Brain-Machine Interface}
\newacronym{ML}{ML}{Machine Learning}
\newacronym{ERP}{ERP}{Event-related Potential}
\newacronym{IMD}{IMD}{Implantable Medical Device}
\newacronym{MCPS}{MCPS}{Medical Cyber-Physical Systems}
\newacronym{SS}{SS}{Spread Spectrum}
\newacronym{DSSS}{DSSS}{Direct-Sequence Spread Spectrum}
\newacronym{FHSS}{FHSS}{Frequency Hopping Spread Spectrum}
\newacronym{SNR}{SNR}{Signal-to-Noise Ratio}
\newacronym{IDS}{IDS}{Intrusion Detection Systems}
\newacronym{OWASP}{OWASP}{Open Web Application Security Project}
\newacronym{CWE}{CWE}{Common Weakness Enumeration}
\newacronym{ACL}{ACL}{Access Control List}
\newacronym{BO}{BO}{Buffer Overflow}
\newacronym{OS}{OS}{Operating System}
\newacronym{NCD}{NCD}{Near Control Device}
\newacronym{RCD}{RCD}{Remote Control Device}
\newacronym{BLE}{BLE}{Bluetooth Low Energy}
\newacronym{BtI}{BtI}{Brain-to-Internet}
\newacronym{BtB}{BtB}{Brain-to-Brain}
\newacronym{B/CI}{B/CI}{Human Brain/Cloud Interface}
\newacronym{IoT}{IoT}{Internet of Things}
\newacronym{FDA}{FDA}{U.S. Food and Drug Administration}
\newacronym{EEG}{EEG}{Electroencephalography}
\newacronym{fMRI}{fMRI}{Functional Magnetic Resonance Imaging}
\newacronym{MEG}{MEG}{Magnetoencephalography}
\newacronym{ECoG}{ECoG}{Electrocorticography}
\newacronym{TMS}{TMS}{Transcranial Magnetic Stimulation}
\newacronym{tES}{tES}{Transcranial Electrical Stimulation}
\newacronym{tFUS}{tFUS}{Transcranial Focused Ultrasound}
\newacronym{DBS}{DBS}{Deep Brain Stimulation}
\newacronym{EMG}{EMG}{Electromyography}
\newacronym{tDCS}{tDCS}{Direct Current Stimulation}
\newacronym{tACS}{tACS}{Alternating Current Stimulation}
\newacronym{SDN}{SDN}{Software-Defined Networking}
\newacronym{NFV}{NFV}{Network Function Virtualisation}
\newacronym{EP}{EP}{Evoked Potential}
\newacronym{VEP}{VEP}{Visual Evoked Potential}
\newacronym{AEP}{AEP}{Auditory Evoked Potential}
\newacronym{ET}{ET}{Essential Tremor}

% Path(s) where your graphic files are
\graphicspath{{Images/}}

\definecolor{ao}{rgb}{0.0, 0.5, 0.0}
\newcommand{\yestick}{{\color{ao}\ding{51}}}
\newcommand{\notick}{{\color{red}\ding{55}}}

% Correct bad hyphenation here
\hyphenation{}

\begin{abstract}
\glspl{BCI} have significantly improved the patients’ quality of life by restoring damaged hearing, sight, and movement capabilities. After evolving their application scenarios, the current trend of BCI is to enable new innovative brain-to-brain and brain-to-the-Internet communication paradigms. This technological advancement generates opportunities for attackers since users' personal information and physical integrity could be under tremendous risk. This work presents the existing versions of the \gls{BCI} life-cycle and homogenizes them in a new approach that overcomes current limitations. After that, we offer a qualitative characterization of the security attacks affecting each phase of the \gls{BCI} cycle to analyze their impacts and countermeasures documented in the literature. Finally, we reflect on lessons learned, highlighting research trends and future challenges concerning security on \glspl{BCI}.
\end{abstract}

%% The code below is generated by the tool at http://dl.acm.org/ccs.cfm.
%% Please copy and paste the code instead of the example below.



%% Keywords. The author(s) should pick words that accurately describe
%% the work being presented. Separate the keywords with commas.

%% This command processes the author and affiliation and title
%% information and builds the first part of the formatted document.
\maketitle

\section{Introduction}
\begin{figure}[ht]
\begin{center}
\includegraphics[width=0.45\textwidth]{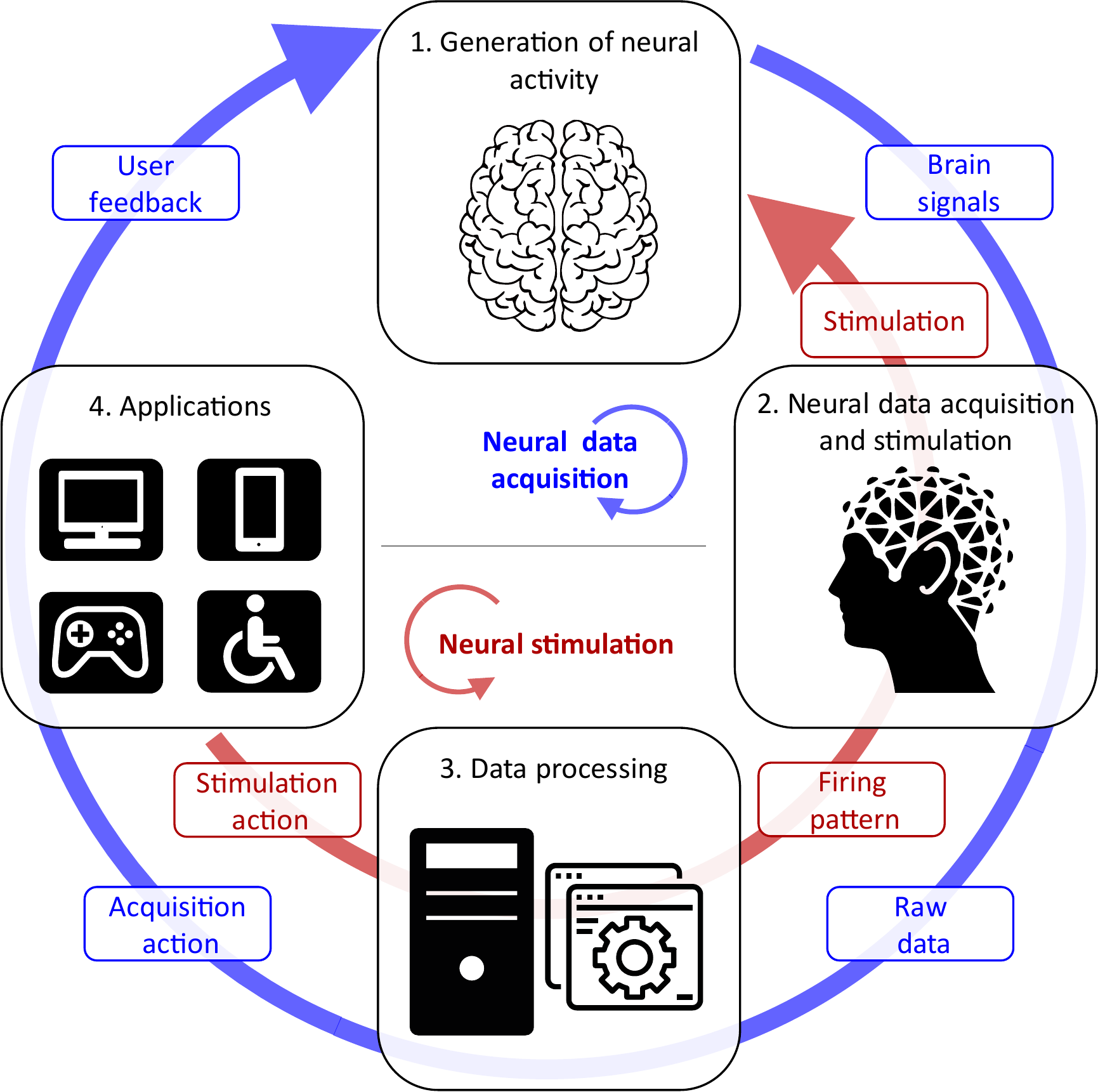}
\end{center}
\caption{General functioning of a bidirectional \gls{BCI}. The clockwise flow indicated with a blue arrow represents the neural data acquisition process, while the counterclockwise flow represented with a red arrow models the brain stimulation.}
\label{fig:intro1}
\end{figure}

\glsreset{BCI}
\label{sec:intro}
% General functioning of BCIs
\glspl{BCI} emerged in the 1970s intending to acquire and process users' brain activity to perform later specific actions over external machines or devices \cite{li:bciApplications:2015}. After several decades of research, this functionality has been extended by enabling not only neural activity recording but also stimulation \cite{tyler:nonInvasiveStimulation:2017}. \figurename~\ref{fig:intro1} describes a simplification of the general components and processes defining a common \gls{BCI} cycle in charge of recording and stimulating neurons \cite{Chizeck:bciAnonymizer:2014, ienca:hackingBrain:2016, ahn:gamesBCIReview:2014}, later presented in Section~\ref{sec:cycle}. It is important to note that these phases are not standard, so we include the most common ones used in the literature. The clockwise direction, indicated in blue, shows the process of acquiring neural data, and the counterclockwise represents the stimulation one, which is highlighted in red. Regarding the neural data acquisition, neurons interact with each other, producing neural activity, either based on previously agreed actions, such as controlling a joystick, or generated spontaneously (phase 1 of \figurename~\ref{fig:intro1}). This activity is acquired by the \gls{BCI} and transformed into digital data (phase 2). After that, data is analyzed by the \gls{BCI} data processing system to infer the action intended by the user (phase 3). Finally, applications execute the intended action, enabling the control of external devices. These applications can present optional feedback to the users, which allows the generation of new neural activity. On the other hand, the counterclockwise direction of \figurename~\ref{fig:intro1} starts in phase 4, where applications define the intended stimulation actions to perform. Phase 3 processes this action to determine a firing pattern containing all the essential parameters required by the \gls{BCI} to stimulate the brain. Finally, the firing pattern is sent to the \gls{BCI}, which is in charge of stimulating specific neurons belonging to one or more brain regions and is dependent on the technology used. In a nutshell, a \gls{BCI} can be a unidirectional or bidirectional communication system between the brain and external computational devices. Unidirectional communications are when they either acquire data or stimulate neurons, while bidirectional communications are when they perform both tasks \cite{rao:bidirectionalBCI:2019}.

% Short security bibliographical review
From the security perspective, \glspl{BCI} are in an early and immature stage. The literature has not considered security a critical aspect of \glspl{BCI} until recent years, where terms such as neurosecurity, neuroprivacy, neuroconfidentiality, brain-hacking, or neuroethics have emerged \cite{denning:neurosecurity:2009, ienca:neuroprivacy:2015, ienca:hackingBrain:2016}. Existing works of the literature have detected specific security attacks affecting \gls{BCI} integrity, confidentiality, availability, and safety, but they do not perform a comprehensive analysis and miss relevant concerns \cite{martinovic:feasibility:2012, bonaci:appStores:2015, li:bciApplications:2015, takabi:privacyThreatsCounter:2016, Sundararajan:privacySecurityIssues:2017}. More specifically, the use of neurostimulation \glspl{BCI} in clinical environments introduces severe vulnerabilities that can have a significant impact on the user's health condition \cite{pycroft:brainjacking:2016}. \glspl{BCI} already existing on the market would benefit from the implementation of robust security solutions, reducing their impact, particularly in clinical environments. Furthermore, the expansion of \glspl{BCI} to new markets, e.g., video games or entertainment, generates considerable risks in terms of data confidentiality \cite{martinovic:feasibility:2012, li:bciApplications:2015, takabi:privacyThreatsCounter:2016, Sundararajan:privacySecurityIssues:2017}. In this context, users' personal information, such as thoughts, emotions, sexual orientation, or religious beliefs, are under threats if security measures are not adopted \cite{martinovic:feasibility:2012, ienca:hackingBrain:2016, takabi:privacyThreatsCounter:2016}. Besides, contemporary \gls{BCI} approaches, such as the use of silicon-based interfaces, introduce new security challenges due to the increase in the volume of acquired data and the use of potentially vulnerable technology \cite{Obaid:silicon-BCI:2020}. The technological revolution of recent years, combined with movements such as the \gls{IoT}, brings an acceleration in the creation of new devices lacking security standards and solutions based on the concepts of \textit{security-by-design} and \textit{privacy-by-design} \cite{bonaci:appStores:2015, takabi:privacyThreatsCounter:2016, Sundararajan:privacySecurityIssues:2017, Ramadan:controlSignalsReview:2017} \cite{ienca:bciConsumerDevices:2018}. This revolution also brings to reality prospective and disruptive scenarios, where we highlight as examples the direct communications between brains, known as \gls{BtB} or Brainets \cite{Pais-Vieira:BtB:2013, zhang:BtB:2019, PaisVieira:brainet:2015, jiang:brainets:2019}, and brains connected to the Internet (\gls{BtI}), which will require significant efforts from the security prism.

% Contribution + aims
Once summarized the functioning of \glspl{BCI} and their security status, the scope of this paper lies in analyzing the security issues of software components that intervene in the processes, working phases, and communications of \glspl{BCI}. Besides, this work considers the security concerns of infrastructures, such as computers, smartphones, and cloud platforms, where different \gls{BCI} architectures are deployed. It is also important to note that, despite this article indicates overall impacts over the brain and the user's physical safety, the main focus of this work is to perform a security analysis from a technological point of view. Aligned with these aspects, and to the best of our knowledge, this article is the first work that exhaustively reviews and analyses the \gls{BCI} field from the security point of view. Since these aspects have not been studied in depth before and \gls{BCI} technologies are still immature, this line of work has a particular interest in a medium to long term. However, this area of knowledge is relevant nowadays, since devices already available on the market need to be protected against attacks.

In this context, Section~\ref{sec:cycle} focuses on analyzing the security issues related to the design of the \gls{BCI} life-cycle. We unify the existing heterogeneous \gls{BCI} life-cycles in a novel and common approach that integrates recording and stimulation processes. Once proposed the new life-cycle design approach, we review the attacks applicable to each phase of the cycle, the impact generated by the attacks and the countermeasures to mitigate them, both documented in the literature and detected by us. After highlighting the security issues related to the \gls{BCI} design, Section~\ref{sec:bciDeployments} reviews the inherent cyberattacks, impacts, and countermeasures affecting current \gls{BCI} deployments scenarios. This section identifies the security issues generated by the devices implementing each life-cycle phase's responsibilities, as well as the communication mechanisms and the application scenarios. The last main contribution of this article is Section \ref{sec:bciTrendChallenges}, where we give our vision regarding the trend of \gls{BCI} and the security challenges that this evolution will generate in the future. Finally, Section~\ref{sec:conclusions} presents some conclusions and future work.

\section{Cyberattacks affecting the BCI cycle, impacts and countermeasures}
\label{sec:cycle}

\begin{figure*}[htbp]
	\centering
	\includegraphics[width=0.95\columnwidth]{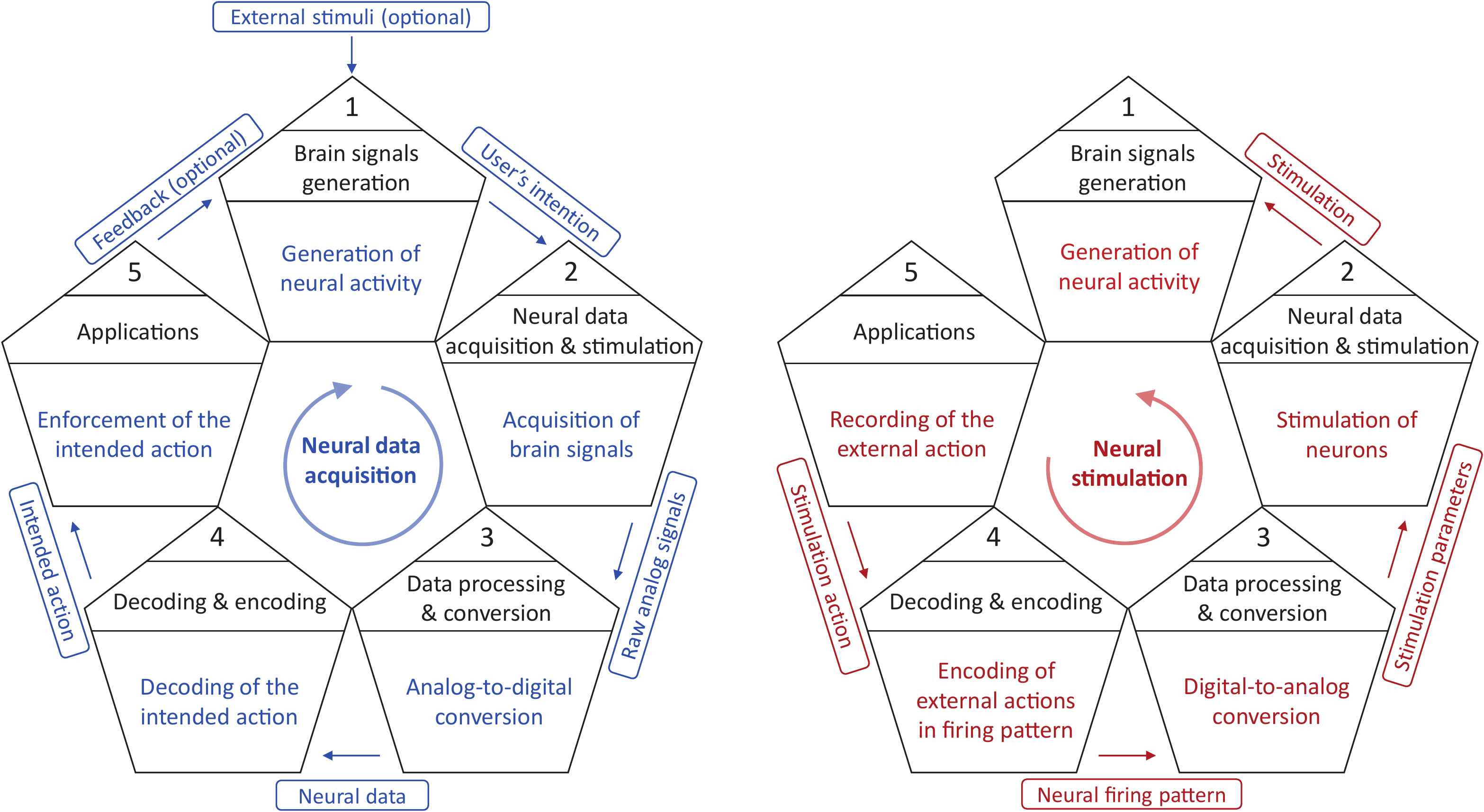}
    \caption{Bidirectional \gls{BCI} functioning cycle representing, in black, the common phases for neural data acquisition and brain stimulation. (Left side) Representation, in blue, of the processes performed and the data transferred by each phase of the neural data acquisition process. This cycle can be seen as a closed-loop process because it starts and ends at the same phase. (Right side) Representation, in red, of the processes and transitions of each phase making up the stimulation process.}
    \label{fig:BCIcycle}
\end{figure*}

This section reviews the different operational phases of \glspl{BCI} detected in the literature, known as the \gls{BCI} cycle, and homogenizes them in a new approach shown in \figurename~\ref{fig:BCIcycle}. After that, we survey the security attacks affecting each phase of the cycle, their impacts, and the countermeasures documented in the literature. We present as well unexplored opportunities in terms of cyberattacks, and countermeasures affecting each phase.

The literature has proposed different configurations of the \gls{BCI} cycle. However, the existing versions only consider the signal acquisition process, missing the stimulation of neurons. These solutions present various classifications of the \gls{BCI} cycle, as some do not consider the generation of brain signals as a phase, or group several phases in only one, without providing information about their roles \cite{Chizeck:bciAnonymizer:2014, ienca:hackingBrain:2016}. Other solutions, as proposed in \cite{vanGerven:bciCycle:2009, li:bciApplications:2015, ienca:hackingBrain:2016, arico:passiveBCI:2018}, are confusing due to they define as new phases, transitions, and data exchanged between different stages. In terms of applications, some authors define a generic stage of applications \cite{ahn:gamesBCIReview:2014, Chizeck:bciAnonymizer:2014, li:bciApplications:2015, Sarma:preprocessingFeature:2016} while others deal with the concept of \textit{commands} sent to external devices \cite{bonaci:appStores:2015, bonaci:exocortex:2015, Sundararajan:privacySecurityIssues:2017, vaid:eegReview:2015, bentabet:synchronousBCI:2016, hong:hybridBCI:2017, chakladar:bookFeatureExtract:2018}. Also, just a few works define the feedback sent by applications to users \cite{vanGerven:bciCycle:2009, vaid:eegReview:2015, li:bciApplications:2015, bonaci:appStores:2015, bonaci:exocortex:2015, Sundararajan:privacySecurityIssues:2017, bentabet:synchronousBCI:2016, ienca:hackingBrain:2016, chakladar:bookFeatureExtract:2018}. To homogenize the \gls{BCI} cycle and address the previously missing or confusing points, we present a new version of the \gls{BCI} cycle with five phases (with clearly defined tasks, inputs, and outputs) that consider both acquisition and stimulation capabilities. \figurename~\ref{fig:BCIcycle} represents our proposal, where the clockwise direction corresponds to the brain signal acquisition process. The information and tasks concerning this functioning are indicated in blue. In contrast, the stimulation process is indicated in the counterclockwise direction, starting from phase 5, and, in each phase, the information and tasks are identified in red. 

According to the neural acquisition process (clockwise direction in \figurename~\ref{fig:BCIcycle}), phase 1 focuses on the generation of brain signals. Generated data contain the user's intention to perform particular tasks; for example, controlling an external device. This phase can be influenced by external stimuli, producing modifications in the regular neural activity. In phase 2, the brain waves are captured by electrodes using a wide variety of technologies, such as \gls{EEG} or \gls{fMRI}. Raw analog signals containing the user's intention are then transmitted to phase 3, where data processing and conversion are required. In particular, this phase performs an analog-to-digital conversion procedure to allow further processing of the data. One of the main goals of this phase is to maximize the \gls{SNR}, which compares the level of the target signal to background noise level to obtain the original signal as accurately as possible. Phase 4 processes the digital neural data to decode the user's intended action, where relevant features are calculated and selected from the neural data. After that, different models (e.g., classifiers, predictors, regressors) or rule-based systems determine the intended action \cite{chakladar:bookFeatureExtract:2018, Sarma:preprocessingFeature:2016}. The action finally arrives at applications in phase 5, which execute the action. Applications can also send optional feedback to the user to generate brain signals and thus new iterations of the cycle. 

Regarding the stimulation process (counterclockwise direction in \figurename~\ref{fig:BCIcycle}), the loop starts in phase 5, where it is specified the stimulation action in a general way (e.g. stimulate a particular brain region to treat Alzheimer's disease). This intended action is transmitted to phase 4, where this input is processed by different techniques, such as \gls{ML}, to generate a firing pattern that contains high-level information about the stimulation devices to be activated, the frequencies used and the temporal planning. Phase 3 intends to transform the firing pattern received, indicated in a general fashion, to specific parameters related to the \gls{BCI} technology used. For example, the identification of neurons to stimulate or the power and voltage required for the process. Phase 2 transmits these stimulation parameters to the stimulation system, that is in charge of the physical stimulation of the brain. After this process, the brain generates neural activity as a response, which can also be acquired by the \gls{BCI} to measure the state of the brain after each stimulation process. At this point, an alternation between brain stimulation and signal acquisition is possible, moving from one direction of \figurename~\ref{fig:BCIcycle} to the other.

Before reviewing the attacks, impacts and countermeasures of each phase of the \gls{BCI} cycle, it is essential to accurately define the concept of \textit{security}, which refers to the "protection of information and information systems from unauthorized access, use, disclosure, disruption, modification, or destruction to provide integrity, confidentiality and availability" \cite{Scholl:security_definition:2008}. The concepts of integrity, confidentiality and availability, together with the concept of \textit{safety}, are used in this section as metrics to evaluate the impact of security attacks against BCI systems. The standard definition of these concepts is the following:

\begin{itemize}
    \item \textbf{Integrity}: "protection against unauthorized modification or destruction of information. A state in which information has remained unaltered from the point it was produced by a source, during transmission, storage, and eventual receipt by the destination". \cite{Kuhn:integrity_definition:2001}
    
    \item \textbf{Confidentiality}: "preservation of authorized restrictions on access and disclosure, including means for protecting personal privacy and proprietary information". \cite{Scholl:security_definition:2008}
    
    \item \textbf{Availability}: "property that data or information is accessible and usable upon demand by an authorized person." \cite{Scholl:security_definition:2008}
    
    \item \textbf{Safety}: "freedom from conditions that can cause death, injury, occupational illness, damage to or loss of equipment or property, or damage to the environment." \cite{Ross:safety_definition:2019}. This work considers the safety concept from the physiological, psychiatric, and psychological perspectives. 
\end{itemize}

At this point, it is essential to note that in this document, the safety concept refers to the preservation of the physical integrity of \gls{BCI} users, not focusing on the conservation of objects or the environment. To better understand the attacks and countermeasures later discussed in this section, \tablename~\ref{table:cycle_attacks} offers a brief description of the attacks affecting \gls{BCI}, whereas \tablename~\ref{table:cycle_countermeasures} describes their countermeasures. For each phase of the \gls{BCI} cycle, we detail the particularities of these attacks and countermeasures.  \\

\renewcommand\arraystretch{0}
\begin{footnotesize}
\setlength\tabcolsep{2pt}
\begin{longtable}{@{}>{\raggedright}p{0.2\linewidth}>{\raggedright\arraybackslash}p{0.78\linewidth}@{}}
\caption{Definition of the attacks detected for the BCI cycle}
\label{table:cycle_attacks} \\

\toprule

Attack
& Description
\\ \midrule

Adversarial attacks \cite{finlayson:adversarialAttacks:2019, liu:adversarialML:2018}
& Presentation of intentionally crafted inputs to an ML system to disrupt its normal functioning and output.
\\\midrule

Misleading stimuli attacks \cite{Landau:security_EEG:2020, martinovic:feasibility:2012, Frank:subliminalProbe:2017}
& Presentation of malicious sensory or motor stimuli to users aiming to generate a specific neural response.
\\\midrule

Buffer Overflow attacks \\ \cite{saito:bufferOverflow:2016, black:bufferOverflow:2016, CWE-119}
& Access to out-of-bounds memory spaces due to insecure software implementations. They take advantage of operations over memory buffers whose boundaries are not well managed.
\\\midrule

Cryptographic attacks \cite{ienca:neuroprivacy:2015, ienca:hackingBrain:2016}
& Exploit vulnerabilities in the elements that define a system, such as algorithms, protocols or tools. A variety of techniques focused on evading the security measures of cryptographic systems.
\\ \midrule

Firmware attacks \cite{vasile:firmwareExtraction:2019, bettayeb:firmware:2019}
& Extract or modify the firmware of a device, a critical piece of software that controls its hardware.
\\ \midrule

Battery drain attacks \cite{Camara:IMDsecurity:2015} \cite{pycroft:securityIMD:2018}
& Consume the battery of a device, reducing its performance or even making it permanently inaccessible.
\\ \midrule

Injection attacks \cite{OWASP:Injection:2017, CWE-74}
& Present an input to an interpreter containing particular elements that can modify how it is parsed, taking advantage of a lack of verification of the input.
\\ \midrule

Malware attacks \\ \cite{chakkaravarthy:malware:2019, yan:mobileMalware:2018, Kurose:computerNetworking:2017}
& Use of hardware, software or firmware aiming to gain access over computational devices to perform malicious actions intentionally.
\\ \midrule

Ransomware attacks \\ \cite{al-rimy:ransomware:2018, fernandezMaimo:ransomwareMCPS:2019}
& Encrypt users' data and demand later an economic ransom to decipher it.
\\ \midrule

Botnet attacks \\ \cite{mahmoud:botnet:2015, amini:botnet:2015}
& Use of botnets, networks of infected devices controlled and coordinated by an attacker, to perform particular attacks directed to specific targets.
\\ \midrule

Sniffing attacks \cite{anu:sniffingAttacks:2017}
& Acquisition of private information by listening to a communication channel. When the data is not encrypted, attackers have access to the content of the whole communication.
\\ \midrule

Man-in-the-middle attacks \cite{Sundararajan:privacySecurityIssues:2017}
& Alteration of the communication between two entities, making the extremes believe that they are communicating directly between each other.
\\ \midrule

Replay attacks \\ \cite{Tanenbaum:computerNetworks:2011, Kurose:computerNetworking:2017}
& Retransmission of previously acquired data to perform a malicious action, such as the impersonation of one of the legitimate participants of the communication.
\\ \midrule

Social engineering attacks \cite{hatfield:socialEngineering:2018, gupta:phishing:2016}
& Psychological manipulation to gain access over restricted resources. An example is phishing attacks, based on the impersonation of a legitimate entity in digital communication.
\\ \midrule

Spoofing attacks \\ \cite{stallings:cryptographySecurity:2017, Tanenbaum:computerNetworks:2011}
& Masquerade an entity of the communication, transmitting malicious data. Frequent spoofing attacks in network communications are, among others, IP spoofing and MAC spoofing.
\\
\bottomrule

\end{longtable}
\end{footnotesize}

\renewcommand\arraystretch{0}
\begin{footnotesize}
\setlength\tabcolsep{2pt}
\begin{longtable}{@{}>{\raggedright}p{0.25\linewidth}>{\raggedright\arraybackslash}p{0.73\linewidth}@{}}
\caption{Definition of the countermeasures detected for the BCI cycle}
\label{table:cycle_countermeasures} \\

\toprule

Countermeasure
& Description
\\ \midrule

Training sessions, demos \\ and serious games \cite{ienca:hackingBrain:2016}
& Initiatives to increase the awareness of the users about the risks of technology.
\\ \midrule

User notifications \cite{Camara:IMDsecurity:2015}
& Alert the users in case an attack is detected, to take part in the defence (e.g. stop using the device).
\\ \midrule

Directional antennas \cite{zou:jamming:2016}
& Antennas that radiate or receive the energy mainly in particular directions, aiming to reduce interference.
\\ \midrule

Analysis of the medium \\ \cite{ienca:hackingBrain:2016}
& Sensing of the communication medium to detect abnormal behavior.
\\ \midrule

Low transmission power \\ \cite{vadlamani:jammingSurvey:2016}
& Reduction of transmission power to avoid the interception of the communication by malicious entities.
\\ \midrule

Frequency and channel \\ hopping \cite{zou:jamming:2016, grover:jamming:2014}
& Wireless communication models based on pseudo-random hopping patterns previously known by sender and receiver.
\\ \midrule

Spread spectrum \\ \cite{vadlamani:jammingSurvey:2016, zou:jamming:2016, Tanenbaum:computerNetworks:2011}
& Transmission of the information in a broader bandwidth to avoid interference in the wireless medium.
\\ \midrule

Access control mechanisms \\ \cite{takabi:privacyThreatsCounter:2016, Takabi:firewallBrain:2016, Camara:IMDsecurity:2015}
& Means of detecting and preventing unauthorized access to particular resources.
\\ \midrule

Privilege management \\ \cite{CWE-120, CWE-121, CWE-122}
& Assign privileges to different groups of users based on roles.
\\ \midrule

Whitelists and blacklists \\ \cite{CWE-77}
& List of entities, such as systems or users, that are allowed or forbidden, respectively, to perform specific actions. 
\\ \midrule

Cryptographic mechanisms \\ \cite{ballarin:cybersecurityAnalysis:2018}
& Use of encryption and decryption techniques to protect the privacy of data, since unprotected information can be accessed and modified by attackers.
\\ \midrule

Differential privacy \cite{liu:adversarialML:2018, ienca:bciConsumerDevices:2018}
& Cryptographic mechanism based on the addition of noise to the data aiming to suppress sensitive aspects, accessible when combined with a large amount of a user's data.
\\ \midrule

Homomorphic encryption \\ \cite{liu:adversarialML:2018}
& Cryptographic mechanism allowing the computation of mathematical operations over ciphered data, generating an encrypted result.
\\ \midrule

Functional encryption \\ \cite{takabi:privacyThreatsCounter:2016, Takabi:firewallBrain:2016}
& Cryptographic mechanism where having a secret key allows to learn a function of encrypted data without revealing the data itself.
\\ \midrule

Authenticity verification \\ \cite{ballarin:cybersecurityAnalysis:2018}
& Ensure that the data we are accessing, or the endpoint we are communicating, is who it claims to be.
\\ \midrule

Legitimacy verification \\ \cite{ballarin:cybersecurityAnalysis:2018}
& Review if a malicious software application has replaced a legitimate one.
\\ \midrule

Feature limitation \cite{OWASP:Misconfiguration:2017}
& Ensure that any software only implements the specific functionality for which it was intended. 
\\ \midrule

Periodic updates \cite{fernandezMaimo:ransomwareMCPS:2019}
& Correct detected vulnerabilities and include new functionalities to reinforce the existing countermeasures.
\\ \midrule

Robust programming \\ languages \cite{CWE-120}
& Choose the most adequate languages taking into consideration their strengths and weaknesses.
\\ \midrule

Compilation techniques \\ and options \cite{CWE-121}
& Specific capabilities of compilers to protect out of bounds accesses to the device memory or CPU registers.
\\ \midrule

Application hardening \\ \cite{haupert:appHardening:2018}
& Modification of an application to make it more resistant against attacks, such as the obfuscation of the application code.
\\ \midrule

Segmented application \\ architectures \cite{saito:bufferOverflow:2016}
& Isolation of architectures and systems, establishing different containers and security groups to communicate with each other.
\\ \midrule

Sandboxing \cite{miramirkhani:sandbox:2018}
& Isolate the execution of different programs, allowing its protection against attacks.
\\ \midrule

Antivirus \cite{stallings:cryptographySecurity:2017}
& Software focused on the prevention, detection, and elimination of malware attacks. Modern antivirus offer protection for a wide variety of threats.
\\ \midrule

Malware visualization \cite{fu:malwareVisualisation:2018}
& Technique focused on the analysis of software binaries in a graphical way to detect anomalous malware patterns.
\\ \midrule

Quarantine of devices \cite{amini:botnet:2015}
& Isolation of infected or potentially infected software, to avoid further propagation and infection.
\\ \midrule

Backup plans \cite{amara:cloudComputing:2017}
& Recurrent copy of data stored in a different location to allow its recovery in case of data loss.
\\ \midrule

Defense distillation \cite{liu:adversarialML:2018}
& Creation of a second ML model based on the original, with less sensitivity regarding input perturbations and offering smoother and more general results.
\\ \midrule

Data sanitisation \cite{Jagielski:adversarialMLregression:2018}
& Rejection of samples that can produce a negative impact on the model, preprocessing and validating all input containing adversarial information.
\\ \midrule

Adversarial training \cite{goodfellow:adversarialML:2018}
& Inclusion of adversarial samples in the training process to allow the recognition of attacks in the future.
\\ \midrule

Monitoring systems \cite{birajdar:forensicImage:2013}
& Capture and analyze the behavior of the entities within a system and their communications. 
\\ \midrule

Anomaly detection \cite{Camara:IMDsecurity:2015}
& Detection of odd behaviors on systems that can potentially correspond to an attack situation.
\\ \midrule

Firewall \cite{stallings:cryptographySecurity:2017}
& Cybersecurity system that only allows incoming or outgoing network communications previously authorized.
\\ \midrule

IDS \cite{stallings:cryptographySecurity:2017}
& Analysis of the network activity to identify potentially damaging communications aiming to disrupt the system.
\\ \midrule

Communication interruption \cite{kirubavathi:botnetAndroid:2018}
& Detention of an active communication to mitigate the impact of an attack if there is evidence of its presence.
\\ \midrule

Input validation \cite{OWASP:Injection:2017}
& Analysis and preprocessing of inputs presented to a system to suppress potential causes of failure.
\\ \midrule

Randomization \cite{takabi:privacyThreatsCounter:2016}
& Change of existing data in a way that does not follow a deterministic pattern and prevents privacy leakage.
\\ \midrule

BCI Anonymizer \cite{bonaci:appStores:2015}
& Anonymization of brain signals acquired from the brain to be shared without exposing users sensitive information.
\\

\bottomrule
\end{longtable}
\end{footnotesize}

\begin{figure*}[htbp]
	\centering
	\includegraphics[width=\columnwidth]{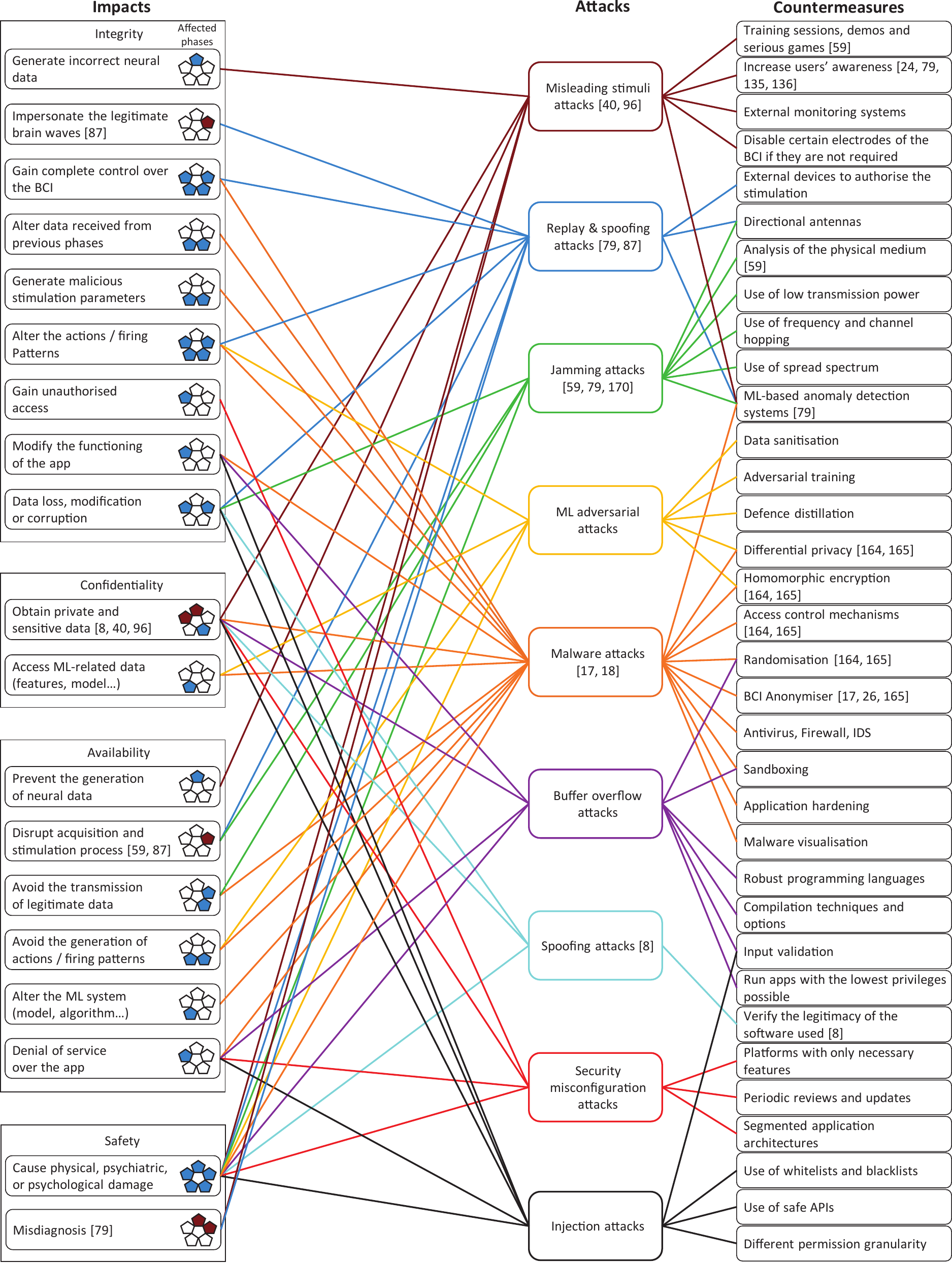}
    \caption{Relationship between the attacks, impacts, and countermeasures over the \gls{BCI} cycle. The phases of the cycle colored in red for each impact represent issues documented in the literature, while those marked in blue are our contribution. The attacks, impacts and countermeasures followed by references have been documented in the literature, and those without a cite represent our contribution.}
    \label{fig:attacksCycle}
\end{figure*}

\figurename~\ref{fig:attacksCycle} indicates the attacks, impacts, and countermeasures described in this section. As can be seen, each attack is represented by a color that associates the impacts it generates and the countermeasures to mitigate it. For each impact included in the figure, it includes a simplified version of the \gls{BCI} cycle. Those phases of the cycle marked in red indicate impacts detected in the literature for that specific phase, whereas the blue color indicates our contribution. Besides, the attacks, impacts and countermeasures marked with references have been proposed in the literature, while those without references are our contribution. It is important to note that this figure highlights the limitations exposed by the literature, as can be appreciated by the volume of our contributions. To simplify the image, we have synthesized most of the safety impacts into a general entry "Cause physical damage", describing the specific impacts over users' health in detail throughout the section.

%%%%%%%%%%
\subsection{Phase 1. Brain signals generation}
\label{subsec:cycleGeneration}

\subsubsection{Attacks}
Considering the neural data acquisition flow, this first phase focuses on the brain processes that generate neural activity, which can be influenced by external stimuli. The literature has detected \textit{misleading stimuli attacks} \cite{Landau:security_EEG:2020, martinovic:feasibility:2012, Frank:subliminalProbe:2017}, a mechanism to alter the brain signals generation by presenting intentionally crafted stimuli to BCI users. To understand these attacks, it is important to introduce some concepts. \textit{\glspl{ERP}} are neurophysiological responses to a cognitive, sensory, or motor stimulus, detected as a pattern of voltage variation \cite{Chizeck:bciAnonymizer:2014}. Within the different types of \glspl{ERP}, \glspl{EP} focus on sensory stimuli and can be divided into two categories, \glspl{VEP} and \glspl{AEP}, related respectively with visual and auditory external stimuli. Specifically, \textit{P300} is a \gls{VEP} detected as an amplitude peak in the \gls{EEG} signal about 300ms after a stimulus, extensively used due to its quick response \cite{sowndhararajan:P300:2018}.  

On the one hand, Martinovic et al. \cite{martinovic:feasibility:2012} used the P300 potential to obtain private information from test subjects and demonstrated misleading stimuli attacks. Visual stimuli were presented in the form of images, grouped as follows: 4-digit PIN codes, bank ATMs and credit cards, the month of birth and photos of people. The objective of the experiment was to prove that users generate a higher peak in the P300 potential when faced with a known stimulus and, therefore, be able to extract private information. The authors used the Emotiv EPOC 14-channel headset \cite{emotivEpoc:2019}, a commercial \gls{BCI} \gls{EEG} device, showing that information leakage, measured in information entropy, was 10\%-20\% of the overall information, and could be increased to approximately 43\%. On the other hand, Frank et al. \cite{Frank:subliminalProbe:2017} demonstrated the possibility of performing subliminal \textit{misleading stimuli attacks}. To perform the experiments, the same \gls{ERP} concept with P300 potentials was used. In this work, the authors showed information hidden within the visual content projected to 29 subjects, in the form of stimuli with a duration of 13.3 milliseconds, imperceptible to the human eye. The study used \gls{EEG} devices of the brands NeuroSky \cite{neuroSky} and Emotiv \cite{emotiv}. We consider that the previous works are relevant to highlight the importance of security in \gls{BCI}, and additional experiments with a higher number of users are required.  

The literature has documented some well-known methods to present stimuli to users and analyze their neural responses \cite{Sundararajan:privacySecurityIssues:2017, bonaci:appStores:2015, martinovic:feasibility:2012}. For example, to study the neural activity generated after a question in a lie detection test \cite{Landau:security_EEG:2020}. Although these methods do not represent attacks themselves, they are an opportunity to develop new misleading stimuli attacks against \glspl{BCI}, defined as:

\begin{itemize}
    \item \textit{Oddball Paradigm}: specific target stimuli, hidden between a sequence of common non-target stimuli, would generate peaks in \gls{ERP}. For example, to differentiate a known face among several unknown ones. 
    \item \textit{Guilty Knowledge Test}: the response generated by familiar stimuli can be differentiated from the generated by unfamiliar elements. This principle has been used for lie detection.
    \item \textit{Priming}: a stimulus can generate an implicit memory effect that later influences other stimuli.
\end{itemize}

Despite the comprehensive study in the literature on \glspl{AEP}, there are no specific works, to the best of our knowledge, describing attacks over auditory stimuli. However, Fukushima et al. \cite{Fukushima:inaudibleFrequencies:2014} described that inaudible high-frequency sounds could affect brain activity. We detect that this scenario generates new opportunities for attackers since the generation of inaudible auditory stimuli does not require close interaction with the victim, helping the attacker to remain undetected.

Regarding neural stimulation, this phase represents the result of the stimulation process within the brain. Based on a lack of literature defining taxonomies of attacks over the brain, we identify two main attack categories during neurostimulation. The first category consists of taking control of the stimulation process to cause neural tissue damage. These attacks may reproduce or worsen the secondary effects often present during the treatment of neurological conditions, such as Parkinson's disease, either by over-stimulation actions or by preventing the treatment. The feasibility of these attacks is supported by \cite{Parastarfeizabadi:closed_loop:2017, Hartmann:parkinson:2019}, who indicated that the adverse effects of neurostimulation are related to the parameters and patterns of the stimulation. Additionally, we identify another modality of attack in this category, based on recreating known neurological conditions if there is an existing neurostimulation device with access to the regions naturally affected by those diseases. As an example, we identify the possibility of recreating neurodegenerative diseases, such as Parkinson's and Alzheimer's diseases, based on a deterioration of cerebral tissue, and epileptic seizures. Although these attacks are nowadays just theoretical \cite{Lopez:disrupt_neural_signaling:2020}, the advance of prospecting BCI technologies like Neuralink \cite{neuralink}, could result in neurostimulation systems that can cover various parts of the brain, thus introducing these threats.

The second category of attacks focuses on inducing an effect or perception in the user. It is well known that neurostimulation can cause multiple psychiatric and psychological impacts, such as mood variations, depression, anxiety, or suicidal thoughts, as later indicated in Section~\ref{subsubsec:phase2_impacts}. An attacker could magnify these effects with malicious stimulation parameters to take advantage of the user. As an example, the attack could aim to reduce the patient's inhibition to ease the extraction of private information. This situation introduces the possibility of \textit{social engineering attacks} to BCI, where the attacker would not require sophisticated social techniques to manipulate its victims psychologically.

\subsubsection{Impacts}
\label{subsubsec:phase2_impacts}
It is important to note that the \textit{misleading stimuli attacks} detailed for this phase have only been conducted against data confidentiality \cite{Landau:security_EEG:2020, Frank:subliminalProbe:2017}, aiming to extract sensitive data from \gls{BCI} users. However, we consider that they can also affect \gls{BCI} integrity, availability, and safety. These stimuli can alter the normal functioning of this phase, generating malicious inputs for the next stages that can derive on disruptions of the service or incorrect actions aiming to cause physical damage to users. Specifically, Landau et al. \cite{Landau:security_EEG:2020} identified that misleading stimuli attacks performed during a medical diagnose, such as a photosensitive epilepsy test in which different visual stimuli are presented, can derive in a misdiagnosis, affecting the users’ safety. We also identify as feasible that malicious stimuli, both perceptible or subliminal, can affect the users’ mood.

From the perspective of neurostimulation, the attacks above can affect users' health differently according to their previously existing diseases, impacting their physical and psychological safety. The issues related to different \gls{BCI} technologies are detailed in Section~\ref{subsec:cycleAcquisitionStimulation}, indicating general impacts over the brain in this phase. \tablename~\ref{table:side_effects} presents the most common side effects during particular neurostimulation therapies. As can be seen, performing an attack during the stimulation process can aggravate or even generate a wide range of negative impacts on BCI patients. Additionally, the authors of \cite{pycroft:brainjacking:2016, pycroft:securityIMD:2018} highlighted common issues to neurological diseases, such as tissue damage, rebound effects, and denial of stimulation (also affecting the service availability). Besides, they identified that an alteration of voltage, frequency, pulse width, or electrode contact used to stimulate the brain could modify the volume of cerebral tissue activated, inducing non-desired effects in the surrounding structures depending on the electrode location and stimulation technique. Pycroft et al. \cite{pycroft:securityIMD:2018} also indicated that an attack on neurostimulation could induce a patient's thoughts and behavior. In \cite{Marin:securityNeurostimulators:2018}, the authors highlighted that attacks on neurostimulation can prevent patients from speaking or moving, cause brain damage or even threaten their life, while the authors of \cite{Landau:security_EEG:2020} indicated the user's frustration if the result of the process is not adequate.

Pycroft et al. \cite{pycroft:brainjacking:2016} indicated potential attacks and harms against neurostimulation patients. First, they detected that an overstimulation procedure could cause tissue damage, independently of the type of stimulation and medical condition. For Parkinson's disease, an attacker could apply a \textasciitilde{}10Hz stimulation over the STN region to produce hypokinesia or akinesia. In patients with essential tremor, where the ventral intermediate nucleus (VIM) is stimulated, both an increase of voltage and a decrease of frequency could dangerously derive in exacerbated tremor. Finally, a variation in the stimulation parameters during the treatment of obsessive-compulsive disorder could generate alterations of reward processing or operant conditioning.

Based on the above, safety impacts are the most damaging in this phase, presenting a risk of irreversible physical and psychiatric issues. In addition, taking advantage of the victim's psychological status, it could ease social engineering attacks as well. The attacker could aim to reduce or inhibit the patient's mental defense mechanisms, acquiring sensitive information, thus impacting data confidentiality. However, more worrisome would be to take advantage of the victim's mental status, in which the patient unconsciously accedes to undesired acts, such as gambling money, buying unnecessary products, committing a crime, or participating in non-consensual sexual intercourse.

\begin{table}[h]
\caption{Summary of the most common side effects during FDA-approved neurostimulation.}
\label{table:side_effects}
\setlength\tabcolsep{2pt}
\renewcommand{\arraystretch}{1}
\resizebox{\textwidth}{!}{
\begin{tabular}{|c|c|c|c|c|}
\hline

\textbf{Technology} & \textbf{Condition} & \textbf{Brain region} & \textbf{Neurological side effects} & \textbf{Psychiatric/psychological side effects} \\ \hline

\multirow{6}{*}{DBS} & \multirow{3}{*}{Parkinson's disease} & STN & \begin{tabular}{@{}l@{}}Akinesia, cramping in the face or hand, \\dysarthria, dysphagia, eyelid apraxia, gait disturbance, \\hypersalivation, impaired vision, incontinence, learning and \\memory difficulties, paresthesia, postural instability, \\speech disturbance, lack of verbal fluency, \\vegetative symptoms, weakness \cite{neuromodulation, Edwards:neurostimulation:2017, Buhman:secondary_effects:2017, Hartmann:parkinson:2019, Dembek:DBS:2017}\end{tabular} & \begin{tabular}{@{}l@{}}Anxiety, apathy, cognitive disturbance, confusion, \\depression, hallucination, submanic state \cite{Edwards:neurostimulation:2017, Buhman:secondary_effects:2017, Hartmann:parkinson:2019}\end{tabular} \\ \cline{3-5} 

    & & GPI & \begin{tabular}{@{}l@{}}Similar to STN \cite{Hartmann:parkinson:2019}\end{tabular} & \begin{tabular}{@{}l@{}}Anxiety, depression, suicidal thoughts \cite{Edwards:neurostimulation:2017, Hartmann:parkinson:2019}\end{tabular}  \\ \cline{3-5} 
    
    & & VIM & \begin{tabular}{@{}l@{}}Dysphagia, fine motor disturbance, \\speech disturbance \cite{neuromodulation}\end{tabular}   &  \\ \cline{2-5} 
    
    & Essential tremor & VIM & \begin{tabular}{@{}l@{}}{}Dysaesthesia, dysarthria, gait disturbance, \\paresthesia, speech disturbance \cite{Edwards:neurostimulation:2017, Buhman:secondary_effects:2017}\end{tabular}   & \\ \cline{2-5} 
    
    & Dystonia & GPI & \begin{tabular}{@{}l@{}}Gait disturbance, paresis, speech disturbance, \\tetanic muscle contractions, visual deficits \cite{Edwards:neurostimulation:2017, Buhman:secondary_effects:2017}\end{tabular} & \begin{tabular}{@{}l@{}}Anxiety, cognitive disturbance, \\confusion, hallucination \cite{Buhman:secondary_effects:2017}\end{tabular}  \\ \cline{2-5}
    
    & \begin{tabular}{@{}c@{}}Obsessive-compulsive \\disorder\end{tabular} & VC/VS, NAc & & \begin{tabular}{@{}l@{}}Depression, operant conditioning, reward processing \\alteration, suicidal thoughts, suicide \cite{Medtronic:OCD:2020} \end{tabular}  \\ \hline
    
RNS & Epilepsy & Seizure origin & \begin{tabular}{@{}l@{}}Death, change in seizures, \\hemorrhage, infection \cite{NeuroPace:manual:2013}\end{tabular} & \begin{tabular}{@{}l@{}}Anxiety, depression, suicide, suicididal thoughts \cite{NeuroPace:manual:2013}\end{tabular} \\ \hline
\end{tabular}}
\end{table}

\subsubsection{Countermeasures}
Focusing on the countermeasures to mitigate misleading stimuli attacks, multiple works \cite{Landau:security_EEG:2020, pycroft:securityIMD:2018, pycroft:brainjacking:2016, Camara:IMDsecurity:2015} identified general measures to raise the awareness of \gls{BCI} users, such as spreading the risks of these technologies among clinicians and patients and the education of the users in these technologies. This is especially interesting since humans usually are the weakest element of a security system. In particular, Ienca et al. \cite{ienca:hackingBrain:2016} indicated that specific training sessions could be beneficial to protect users against potentially unsafe stimuli related to authentication methods and banking-related information. Besides, the inclusion of demos and serious games in commercial \gls{BCI} devices may educate them on the risks of these technologies. However, these countermeasures can only be applied when the user is aware of the stimuli. Because of that, we consider that \textit{misleading stimuli attacks} can be reduced if \glspl{BCI} are complemented with external systems that monitor the stimuli presented and give users the possibility to evaluate if the content is appropriate. For example, by analyzing if the multimedia contents showed to users, such as images or videos, have been maliciously modified \cite{wahab:forensicVideo:2014, birajdar:forensicImage:2013}, even if they are subliminal. Additionally, we propose using predictive models based on anomaly detection systems, aiming to detect an attack in its early stage and deploy mechanisms to mitigate them.

\subsection{Phase 2. Neural data acquisition \& stimulation}
\label{subsec:cycleAcquisitionStimulation}

\subsubsection{Attacks}
This second phase focuses on the interaction of \gls{BCI} devices with the brain to acquire neural data or perform its stimulation. Regarding data acquisition, the authors of \cite{li:bciApplications:2015, Landau:security_EEG:2020} identified the use of a combination of \textit{replay and spoofing attacks} in which previous signals from the \gls{BCI} user, signals from other users, or synthetic signals can impersonate the legitimate brain waves. We detect the applicability of these attacks to stimulation systems, where an attacker can force specific stimulation behaviors based on previous actions. One possible outcome of this control can be an increase in the voltage delivered to the patient's brain \cite{Marin:securityNeurostimulators:2018}. Besides, the authors of \cite{ienca:hackingBrain:2016, Landau:security_EEG:2020} detected the use of \textit{jamming attacks} against the neural data acquisition process, transmitting electromagnetic noise to the medium. Based on Vadlamani et al. \cite{vadlamani:jammingSurvey:2016}, we also identify this problem in neural stimulation, where \textit{jamming attacks} can override the legitimate signals emitted by the \gls{BCI} electrodes if they are transmitted with enough power.

\subsubsection{Impacts}
Regarding the impacts produced by the previous attacks, Li et al. \cite{li:bciApplications:2015} identified that \textit{replay and spoofing attacks} affect both data integrity and availability, being able to disrupt the acquisition process. Landau et al. \cite{Landau:security_EEG:2020} highlighted that these attacks could interfere with clinical diagnosis procedures, replacing the legitimate brain signals by malicious ones, concluding in misdiagnosis, and producing either an absence of treatment or an unnecessary one on healthy patients. We identify that these attacks, applied to the stimulation scenario, can disrupt the stimulation process or acquire and modify the stimulation pattern used by the BCI to maliciously stimulate the neurons, affecting data integrity, data and service availability, and the patient's safety. Focusing on \textit{jamming attacks}, an attacker can aim to prevent the electrodes from capturing brain signals due to the noise transmitted \cite{ienca:hackingBrain:2016, Landau:security_EEG:2020}, affecting their availability and safety. We detect that jamming attacks can also affect neurostimulation scenarios, where signals with enough power can override the legitimate ones, affecting the integrity and availability of the data, as well as the patient's safety during stimulation actions.

Apart from the impacts derived from the previous attacks, it is important to note that each specific \gls{BCI} technology presents specific risks according to their invasiveness and functioning, and thus the impact generated by an attack differs. To analyze this situation, we select some of the most used \gls{BCI} technologies used to acquire neural data or stimulate the brain. For each one of them, we address specific considerations to evaluate their impact.

Regarding the issues related to acquisition technologies, it is necessary to consider both their temporal and spatial resolutions. We identify that a low temporal resolution in acquisition technologies presents concerns on data and service availability since the devices transmit a reduced amount of data that can be affected more easily by electromagnetic interference and, especially, \textit{jamming attacks}. Besides, this situation can also be beneficial for \textit{replay and spoofing attacks}, since attackers have more time to prepare and send malicious data. A high spatial resolution can impact on data confidentiality, allowing attackers to have access to more sensitive neural data. It is worthy to note that attacks on technologies such as \gls{fMRI} or \gls{MEG} can potentially have a higher economic impact due to the high cost of these technologies compared to others like \gls{EEG} \cite{Ramadan:controlSignalsReview:2017, Lebedev:BrainMachineIF:2017}. Nevertheless, \gls{EEG} is the most studied acquisition technology from the security perspective, due to its wide availability outside clinical environments, highlighting the feasibility of attacks such as \textit{misleading stimuli attacks} or \textit{jamming attacks}.

Although the literature has documented some potential security impacts for acquisition technologies, the impact of neurostimulation technologies on patient's health has been studied in a more detailed way, specifically in the field of \glspl{IMD}. Because of that, we first introduce the most common stimulation technologies nowadays to review their specific impact later, mainly addressing safety issues.

Focusing on the specific impacts of neurostimulation technologies, \gls{DBS} is the most studied one due to its invasiveness, where Medtronic is one of the most popular brands commercializing open-loop DBS devices \cite{Parastarfeizabadi:closed_loop:2017}. The side effects of this method have been extensively studied in the literature, where some of them have previously been presented in \tablename~\ref{table:side_effects} for the treatment of particular conditions. According to Pycroft et al. \cite{pycroft:brainjacking:2016}, the use of \gls{DBS} with high charge densities can cause tissue damage. Furthermore, an increase or decrease in the stimulation frequency can have a considerable impact on its efficacy, even reversing the stimulation effect. Finally, an alteration of emotion and affect processing can occur during \gls{DBS} as side-effects, such as pathological crying or inappropriate laughter, having a distressing impact.

Moving to \gls{TMS}, Polan\'ia et al. \cite{polania:modifyingBrainNonInvasive:2018} indicated that pulses applied to particular areas could induce suppression of visual perception or speech arrest, which serves as an opportunity for attackers. León et al. \cite{Leon:TMS:2018} highlighted that \gls{TMS} could produce side-effects such as headache and neck pain, being epileptic seizures possible but improbable. The side effects of \gls{tES} usually are mild, such as skin tingling, itching, and redness \cite{Moreno-Duarte:tES:2014}. Nevertheless, this technique can have indirect effects on the stimulation of non-neuronal elements, such as peripheral nerves, cranial nerves, or retina. Because of that, the stimulation is limited to maximum tolerable doses \cite{Liu:tES:2018}. Besides, in patients with depression, \gls{tDCS} can derive to mania and hypomania cases \cite{Matsumoto:tES:2017}. It is worthy to note that the side effects described above can naturally arise in controlled environments where clinicians have strict control over the procedure. However, if attackers alter the therapy, they could recreate or amplify malicious conditions, generating a clear impact on patients' health.

The Neuropace RNS is a closed-loop neurostimulation system for treating drug-resistant epilepsy, performing both neural data acquisition and neurostimulation procedures. It presents the advantage of delivering stimulation only when detecting the beginning of seizure activity, reducing secondary effects. Nevertheless, it introduces potential challenges than can be used by an attacker to impact its users' safety \cite{Parastarfeizabadi:closed_loop:2017}. First, we identify that the closed-loop behavior could induce, in both clinicians and patients, a reduction of the perception of risks, assuming that the device is working correctly. Furthermore, since the device presents autonomous capabilities, an attacker could disrupt its behavior, without the knowledge of the user, to generate an impact on data confidentiality, service availability, and safety.

\subsubsection{Countermeasures}
Regarding the countermeasures to detect and mitigate replay and spoofing attacks, Landau et al. \cite{Landau:security_EEG:2020} proposed, for data acquisition, the use of anomaly detection mechanisms to detect modified inputs, as well as the accuracy improvement of acquisition devices. Besides, we propose a mechanism able to disable the electrodes not required for the current application usage and avoid potential risks, such as the acquisition of P300 in brain signals. This action could be performed automatically by the \gls{BCI} system or based on the patient's or clinician's decision. Taking into account neural stimulation, and specifically for IMDs, external devices to authenticate and authorize the stimulation actions can be used \cite{Camara:IMDsecurity:2015}. The authors of \cite{grover:jamming:2014, zou:jamming:2016, vadlamani:jammingSurvey:2016} documented several detection mechanisms and countermeasures related to the mitigation of jamming attacks. All detection procedures are based on an analysis of the medium to detect abnormal behavior, as identified for neural data acquisition by Ienca et al. \cite{ienca:hackingBrain:2016}. Specifically, Landau et al. \cite{Landau:security_EEG:2020} proposed using an ensemble of classifiers to detect the addition of noise to the benign input. As proposed countermeasures, Vadlamani et al. \cite{vadlamani:jammingSurvey:2016} identified the use of low transmission power as a possible solution to harden the detection of the legitimate transmission, and the use of directional antennas oriented to the brain to avoid the jamming. The use of frequency hopping \cite{zou:jamming:2016} and channel hopping \cite{grover:jamming:2014} after a particular duration of time also aim to reduce the impact of these attacks. We detect that the use of directional antennas is also a possible solution for \textit{replay and spoofing attacks}. Finally, it is worthy to note that the mitigation of the previous impacts focused on user's safety is the consequence of mitigating the attacks spotted against \gls{BCI} devices.

In the scenario of closed-loop neurostimulation systems, we identify as essential to have information about the behavior of the device, from both acquisition and stimulation procedures. These feedback mechanisms would allow to externally analyze the status of the brain and the stimulation decisions. Another proposal is the use of anomaly detection systems, included in the device, to identify unusual stimulation parameters, or an absence of treatment when a seizure occurs, notifying the user. This second approach could be more energy preserving, and the election of the strategy would depend on the use case.

\subsection{Phase 3. Data processing \& conversion}
\label{subsec:cycleProcessingConversion}

\subsubsection{Attacks}
This phase performs the data processing and conversion tasks required to allow neural data and stimulation actions to be ready for subsequent stages. Although the literature has not detected security problems in this phase, according to the aspects indicated by Bonaci et al. in \cite{bonaci:appStores:2015, bonaci:exocortex:2015}, we identify \textit{malware attacks} as possible against this phase, taking control over the \gls{BCI}. These attacks are candidates to affect both acquisition and stimulation processes, impacting the tasks performed in this phase. In particular, we identify that malware can disrupt the analog-to-digital conversion that occurs during neural data acquisition, as well as the translation of firing patterns to particular stimulation devices. We also detect that \textit{jamming attacks} applied to the previous phase for data acquisition can impact this phase since a distorted input signal with enough noise can be difficult to filter and thus propagate this signal to subsequent phases.

\subsubsection{Impacts}
In this context, we identify that \textit{malware attacks} have an impact on both neural data acquisition and stimulation, where attackers alter or override the data received from previous phases, generating malicious data sent to subsequent phases. That is, the analog data recorded during neural data acquisition or the firing pattern used in neurostimulation processes. These attacks can gather the sensitive data managed in this phase, both analog and digital, and send it to the attackers, affecting data confidentiality. For example, information about private thoughts or neurological treatments. In terms of data and service availability, both acquisition and stimulation are potentially vulnerable to malware that avoids data transmission to subsequent phases of the cycle. Malware affecting integrity and availability is also a threat against users' physical safety, generating damaging stimulation patterns or dangerous actions sent to applications. Besides, the impacts and countermeasures described in the first phase of the acquisition flow for jamming attacks are also applicable to the current stage.

\subsubsection{Countermeasures}
Regarding the countermeasures to mitigate attacks affecting data confidentiality, Chizeck et al. \cite{Chizeck:bciAnonymizer:2014} defined a US patent application entitled ``Brain-Computer Interface Anonymize'' that proposes a technology capable of processing neural signals to eliminate all non-essential private information \cite{bonaci:appStores:2015,takabi:privacyThreatsCounter:2016}. As a result, sensitive information is never stored in the \gls{BCI} device or transmitted outside. We identify this method as especially relevant in this phase, as it is the first stage after the \gls{BCI}'s acquisition process. Although the authors do not provide details about techniques or algorithms to understand how raw signals are processed, they indicate that this process can only be performed on hardware or software within the device itself, and not on external networks or computer platforms, as a way to ensure the privacy of the information. Besides, Ienca et al. \cite{ienca:bciConsumerDevices:2018} proposed the use of \textit{differential privacy} to improve the security and transparency of data processing.

The countermeasures to mitigate malware depend on their type and behavior. We consider the use of antivirus software and \gls{IDS} as alternatives for the protection of individual devices, based on \cite{Landau:security_EEG:2020}. Besides, the authors of \cite{stallings:cryptographySecurity:2017, yan:mobileMalware:2018} considered perimeter security mechanisms, such as \textit{firewalls}, responsible for analyzing all incoming and outgoing communication of the device. We also propose using \gls{ML} anomaly detection systems to identify potential malware threats \cite{Camara:IMDsecurity:2015} \cite{rathore:multilayer:2018}. Finally, Chakkaravarty et al. \cite{chakkaravarthy:malware:2019} reviewed current persistent malware techniques able to bypass common countermeasures and proposed mitigation techniques, such as \textit{sandboxing} \cite{miramirkhani:sandbox:2018}, \textit{application hardening} \cite{haupert:appHardening:2018} and \textit{malware visualization} \cite{fu:malwareVisualisation:2018}. It is essential to highlight that the countermeasures applicable for this phase highly depend on the device constraints that implement this phase, which is typically the \gls{BCI} device (see Section \ref{sec:bciDeployments}).

\subsection{Phase 4. Decoding \& encoding}
\label{subsec:cycleDecodingEncoding}

\subsubsection{Attacks}
\textit{Decoding \& encoding} is the phase focused on identifying the action intended by the users in neural data acquisition or the specification of the neural firing pattern in neurostimulation. \textit{Malware} attacks have been identified in the literature by Bonaci et al. \cite{bonaci:appStores:2015, bonaci:exocortex:2015} from the signal acquisition perspective. Specifically, they identified that attackers could use \textit{malware} to either override the functioning of this phase or to implement additional malicious algorithms. Besides, we identify that \textit{malware} attacks can also be applied to the stimulation flow, avoiding or disrupting a firing pattern's generation. Besides, we identify that \textit{adversarial attacks} can also be applied to this phase for both acquisition and stimulation tasks, taking advantage of the classification algorithms used. These attacks affect all types of \gls{ML} models, and, because of that, they are currently an open challenge \cite{finlayson:adversarialAttacks:2019}. Liu et al. \cite{liu:adversarialML:2018} detected the possibility of \textit{poisoning attacks}, where attackers introduce crafted adversarial samples to the data, aiming to change its distribution. \textit{Evasion attacks} aim to create samples that evade detection systems, whereas \textit{impersonate attacks} focus on adversarial samples that derive in incorrect classification of the legitimate ones. Finally, two attack models exist according to the knowledge about the model \cite{goodfellow:adversarialML:2018}. In \textit{white-box attacks}, adversaries know the model, while in \textit{black-box attacks}, they only have access to the model through a limited interface. 

\subsubsection{Impacts}
The previously described attacks generate particular impacts on \gls{BCI}. On the one hand, \textit{malware} has an impact on data integrity and availability, as it can alter or ignore the received data from previous phases, and override the output of the current one. That is, disrupt the intended action sent to \gls{BCI} applications in the acquisition process, such as preventing the control of a wheelchair or changing its direction, or the firing pattern in neural stimulation, enabling a wide variety of attacks as described in Section~\ref{subsec:cycleGeneration}. Besides, \textit{malware} affects the availability of the \gls{ML} process by the alteration of the trained model or the \gls{ML} algorithm. From a data confidentiality perspective, \textit{malware} can access the features used in the \gls{ML} training phase, as well as gather information about the model and the algorithm used. \textit{Malware} also affects users' safety, as the previous integrity and availability impacts derive in malicious actions and firing patterns that affect the integrity of users, such as causing neural damage or inducing particular psychological states. On the other hand, \textit{adversarial attacks} also affect data integrity and availability, as the introduction of malicious samples aiming to disrupt the model can alter or avoid the generation of actions and firing patterns. Shokri et al. \cite{shokri:inferenceAttacksML:2017} demonstrated that \gls{ML} models are sensitive against \textit{adversarial attacks}, aiming to detect if a sample exists in the model's training dataset. Based on that, an attacker may extract sensitive users' data, such as previous intended actions or used patterns during stimulation actions. Taking into account data confidentiality, Landau et al. \cite{Landau:security_EEG:2020} detected that a malicious entity taking control of the output of this phase could access the user's intention. Finally, the use of malicious samples, as is the case of \textit{poisoning attacks}, alter the \gls{ML} system, deriving in safety impacts for both cycle directions. 

\subsubsection{Countermeasures}
To mitigate the attacks on the \gls{ML} training phase affecting integrity and availability, we have identified several techniques proposed in the literature for generic \textit{adversarial attacks}, that can serve as an opportunity to improve the security of \gls{BCI}. First, \textit{data sanitization} is useful to reject samples containing adversarial information, thus disrupting the model. Jagielski et al. \cite{Jagielski:adversarialMLregression:2018} proposed a similar approach against poisoning attacks applied to regression techniques, where noise and \textit{outliers} are suppressed from the training dataset. Nevertheless, it does not prevent attackers from crafting samples similar to those generated by the legitimate distribution. Countermeasures such as \textit{adversarial training} or \textit{defense distillation} have been presented in this context. However, both have limitations, as they depend on the samples used during the training and can be broken using \textit{black-box attacks} and computationally expensive attacks based on iterative optimization \cite{goodfellow:adversarialML:2018, liu:adversarialML:2018}. Goodfellow et al. \cite{goodfellow:adversarialML:2018} also proposed \textit{architecture modifications}, based on the improvement of \gls{ML} models to be more robust, but this derives in models difficult to train that have degradation in the performance when used in non-adversarial situations. Liu et al. \cite{liu:adversarialML:2018} documented the integration of techniques to mitigate the attacks, called \textit{ensemble method}. They also indicated two methods that can apply in both training and testing phases: \textit{differential privacy} and \textit{homomorphic encryption} \cite{liu:adversarialML:2018, takabi:privacyThreatsCounter:2016, huertas:databaseSecurity:2016}. Finally, it is worthy to note that the countermeasures to mitigate \textit{malware attacks} in the previous phase can apply to the current one.

\subsection{Phase 5. Applications}
\label{subsec:cycleBCIapps}

\subsubsection{Attacks}
From the data acquisition context, applications perform in the physical world the actions intended by users through their neural activity. These actions can range from the interaction with a computer or smartphone, to the control of a robotic limb. From the perspective of neural stimulation, applications are the entry point of the information transmitted to the brain, like sensory stimuli in prosthesis or cognitive enhancement. In this section, we consider attacks on applications, without analyzing their communication with external systems, addressed in Section \ref{subsec:archLocalBCI}. 

Considering the issues of this phase, \textit{spoofing attacks} over \glspl{BCI} have been detected in the literature, where an attacker creates malicious applications identical to the original and make them available in app stores \cite{ballarin:cybersecurityAnalysis:2018}. 
The authors of \cite{li:bciApplications:2015, bonaci:appStores:2015, bonaci:exocortex:2015} identified \textit{malware attacks} as a threat in \gls{BCI}. Besides, Pycroft et al. \cite{pycroft:brainjacking:2016} identified that the use of consumer devices, such as smartphones, generates new risks and security problems. Specific considerations about malware are the same as detailed in Sections \ref{subsec:cycleProcessingConversion} and \ref{subsec:cycleDecodingEncoding}. Moreover, we have found several opportunities related to cyberattacks performed against applications. In particular, we detect security misconfiguration issues, \gls{BO} attacks, and injection attacks over applications. However, the detailed analysis of these particular attacks is out of the scope of this work, and we only address general aspects related to \gls{BCI}.

\subsubsection{Impacts}

Landau et al. \cite{Landau:security_EEG:2020} identified multiple risks on \gls{BCI} applications with the independence of any attack. They detected that an attacker could interfere with the user's ability to use the device, impacting its availability. They also detected confidentiality concerns regarding the identification of users by their neural data, illustrating a scenario in which an attacker extracts \gls{EEG} data from the application and compares it with the \gls{EEG} database of a hospital, identifying the user and accessing his or her medical records. This identification can derive in a discrimination situation based on the belonging of specific groups, such as religious beliefs. Besides, most \gls{BCI} development APIs offer full access over the information and do not implement limitations on the stimuli presented to users, generating confidentiality issues \cite{martinovic:feasibility:2012,Frank:subliminalProbe:2017,bonaci:appStores:2015,li:bciApplications:2015,takabi:privacyThreatsCounter:2016,Sundararajan:privacySecurityIssues:2017}. Finally, all the attacks affecting this phase can force applications to send malicious stimuli or actions, causing physical harm \cite{ballarin:cybersecurityAnalysis:2018}.

Considering the impact of the previous attacks, applications created by \textit{spoofing attacks} affect both data integrity and confidentiality, as they can present malicious stimuli to obtain sensitive neural information, such as thoughts or beliefs \cite{ballarin:cybersecurityAnalysis:2018}. In neurostimulation scenarios, we identify that these fraudulent applications could entirely modify the firing patterns used to stimulate the patient, generating a high impact over safety. More particularly, these applications could induce psychological states in the victim, making them more willing to gamble, or even generate adverse effects such as anxiety and depression. Based on that, the attacker could take advantage of these mental states, injecting in-app advertisements to earn money from the victim.

\textit{Malware attacks} impact the integrity of the applications by altering their services and capabilities, such as disabling the encryption of information. Besides, they can compromise applications' confidentiality, gaining access to sensitive information such as medical records and user profiles used during neurostimulation treatments. Concerning the availability of the application, \textit{malware attacks} can derive in denial of service over the application, impacting in processes such as controlling prosthetic limbs or wheelchairs. 

We detect that \textit{misconfiguration attacks} present data integrity issues, where attackers take advantage of the system to gain unauthorized access, such as weak access control mechanisms. Data confidentiality issues are also present, for example, on configuration files that have static predefined passwords, allowing attackers to gain access to users' private data. Applications' availability problems are also possible, as a misconfiguration issue can serve as a first step to disrupt the normal behavior of the BCI application.

Moving to \textit{injection attacks}, they can produce data loss, modification, and corruption, affecting the integrity of applications \cite{OWASP:Injection:2017, CWE-74}. In terms of confidentiality, they can produce the disclosure of sensitive information to unauthorized parties \cite{OWASP:Injection:2017, CWE-74}, such as insurance companies aiming to select the best candidates for their products \cite{ballarin:cybersecurityAnalysis:2018}. Availability can be affected by a denial of access over an authentication system, or producing crash, exit or restart actions on the applications, disrupting vital processes such as clinical neurostimulation \cite{OWASP:Injection:2017, CWE-78}.

\textit{\gls{BO} attacks} can derive in the execution of unauthorized code or commands, where an attacker can alter the normal functioning of the application or access to sensitive information \cite{CWE-120}. Furthermore, they can also aim to bypass protection mechanisms by the execution of code outside the scope of the program's security policy. These actions can affect the data integrity, confidentiality, and availability of the application \cite{CWE-121}. 

\subsubsection{Countermeasures}

It is necessary to verify the legitimacy of the applications and ensure sufficient control of the app stores to mitigate \textit{spoofing attacks} \cite{ballarin:cybersecurityAnalysis:2018}. In that regard, Landau et al. \cite{Landau:security_EEG:2020} proposed the use of applications developed by authorized organizations to ensure their trustworthiness. When it comes to \textit{malware attacks}, the same countermeasures proposed for the \textit{Data processing \& conversion} phase also apply for applications. That is, the use of antivirus, firewall, \gls{IDS}, and anomaly detection systems to identify and mitigate the attacks. Furthermore, Takabi et al. \cite{takabi:privacyThreatsCounter:2016, Takabi:firewallBrain:2016} proposed the use of access control mechanisms over the information to restrict its access and thus mitigate confidentiality impacts. They also indicated the use of randomization and differential privacy. Besides, they proposed the integration of \textit{homomorphic encryption} to operate with encrypted information combined with \textit{functional encryption} to access only to a subset of the information. 

As an opportunity for \gls{BCI}, we identify some preventive actions against \textit{misconfiguration attacks} defined by the \gls{OWASP} \cite{OWASP:Misconfiguration:2017}, such as the use of minimal platforms with only necessary features, components, libraries, and software to reduce the probability of misconfiguration issues. Moreover, a periodic review and update of configuration parameters are also beneficial as part of the management process of applications. It is also necessary to create segmented application architectures that offer a division between components and defines different security groups, using \glspl{ACL}. 

Concerning \textit{\gls{BO}}, it is important to use programming languages that protect against these attacks, as well as the use of compilers with detection mechanisms. \cite{saito:bufferOverflow:2016}. Developers must validate all inputs and follow well practice rules when using memory (e.g., verification of the boundaries of buffers). Moreover, sensitive applications must be ran using the lowest privileges possible and even isolated using sandbox techniques \cite{CWE-120, CWE-121, CWE-122}. To detect \textit{injection attacks}, both static and dynamic analysis of applications' source code have been proposed \cite{OWASP:Injection:2017}. For their mitigation, it is necessary to escape all special characters included in the input \cite{OWASP:Injection:2017, CWE-78}. Multiple solutions have been proposed, such as the use of whitelists and blacklists \cite{CWE-77}, the use of safe languages and APIs containing automatic detection mechanisms \cite{OWASP:Injection:2017, CWE-74}, the use of sandboxing techniques to define strict boundaries between processes \cite{CWE-78}, the definition of different permissions on the system \cite{CWE-77}, and error messages with minimal but descriptive details.

\section{Security issues affecting the BCI deployments}
\label{sec:bciDeployments}

\begin{figure*}[h]
\begin{center}
\includegraphics[width=0.75\textwidth]{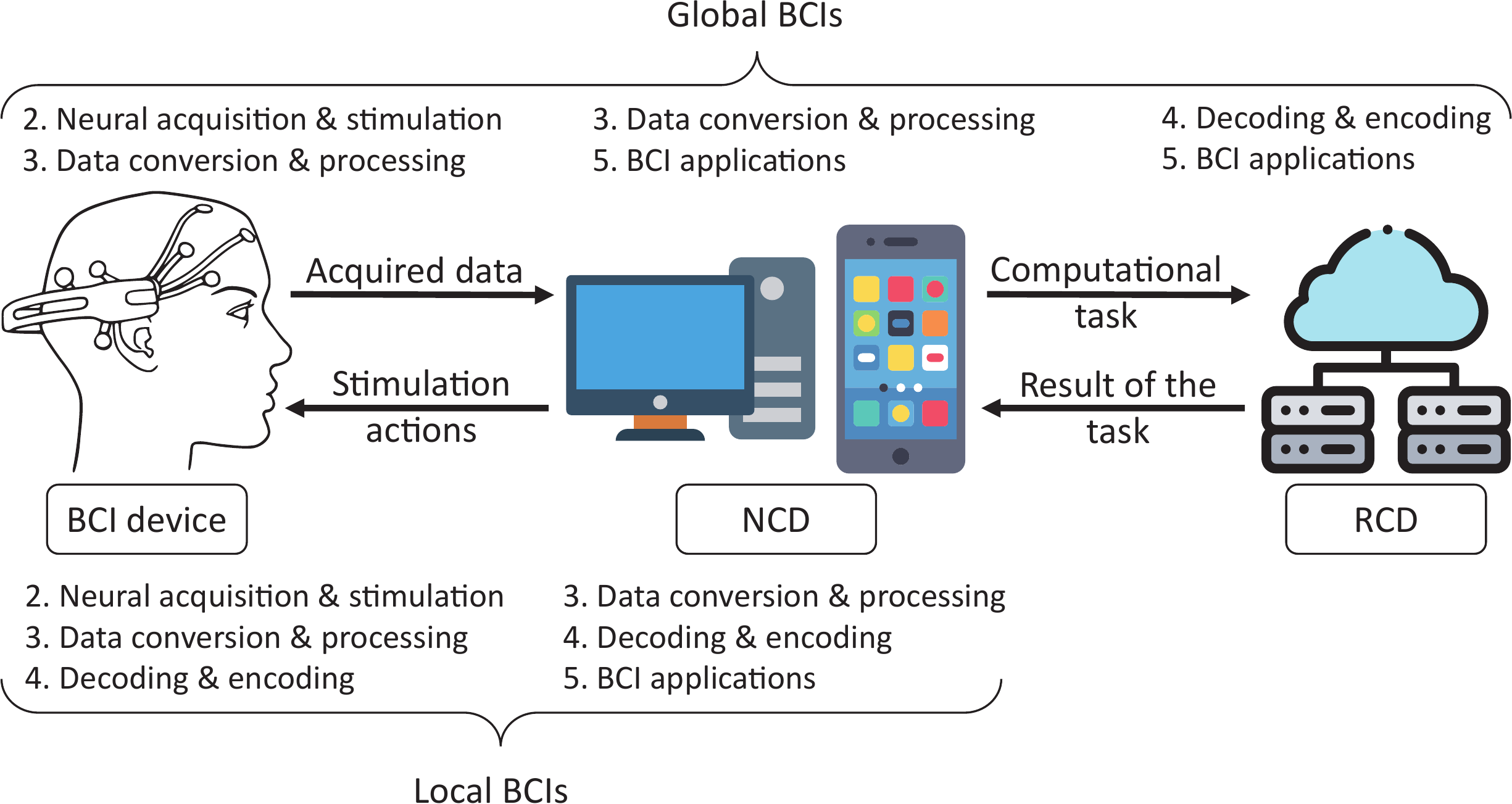}
\end{center}
\caption{Representation of Local \gls{BCI} and Global \gls{BCI} deployments, indicating the communication between their elements and the stages of the \gls{BCI} cycle that each element implements according to the architectural deployment.}
\label{fig:archBCI}
\end{figure*}

This section reviews the different architectural deployments of the \gls{BCI} cycle found in the literature. After that, we group them into two main families, characterized by the \gls{BCI} cycle implementation and its application scenario. In contrast to Section \ref{sec:cycle}, where the security analysis is independent of the deployment, this section reviews the state of the art of existing attacks affecting the devices implementing each phase of the \gls{BCI} cycle, as well as their impacts and countermeasures. New opportunities, in terms of attacks and countermeasures, missed by the literature, are also highlighted in this section. \figurename~\ref{fig:archBCI} represents both architectural deployments defined, Local \glspl{BCI}, and Global \glspl{BCI}, indicating the communication between their elements and the phases of the \gls{BCI} cycle that each element implements according to the type of deployment. 

\subsection{Local \gls{BCI}}
\label{subsec:archLocalBCI}

% Description of the architecture
\subsubsection{Architecture description}
Local \gls{BCI} deployments highlight by managing the neural data acquisition and stimulation processes of single users. This architecture typically deploys the \gls{BCI} phases between two physical devices, as represented in \figurename~\ref{fig:archBCI}. The first one, identified as \textit{\gls{BCI} device}, focuses on the neural acquisition and stimulation procedures (phases 1 and 2 of the \gls{BCI} cycle). In contrast, \gls{BCI} applications (phase 5) run in a \gls{NCD}, a PC or smartphone that controls the \gls{BCI} device using either a wired or wireless communication link. Phases 3 and 4 of the cycle can be implemented equally in both devices, where manufacturers make the final decision. At this point, it is essential to note that alternative designs can arise due to specific requirements of the deployments, such as the presence of multiple users. Moreover, we consider fully implantable \glspl{BCI} within this architecture, since they require an external device for its configuration and verification.

% Examples of deployments
\subsubsection{Examples of deployments}
This kind of architectural deployment is the most commonly implemented for consumer-grade \glspl{BCI}, where commercial brands like NeuroSky or Emotiv focus on scenarios such as gaming and entertainment \cite{martinovic:feasibility:2012, mcmahon:gamesBCI-VR:2018, ahn:gamesBCIReview:2014}. Neuromedical scenarios also use this approach, where an \gls{NCD} placed in the clinical environment manages the acquisition and stimulation processes. This section specifically addresses the issues detected in physical \gls{BCI} devices, the inherent problems of the \gls{NCD}, and those related to the communication between \gls{BCI} and \gls{NCD}. At this point, it is important to note that the attacks, impacts, and countermeasures detected for the \gls{BCI} cycle are also applicable.

% Attacks
\subsubsection{Attacks}
Focusing on \gls{BCI} devices, Ballarin et al. \cite{ballarin:cybersecurityAnalysis:2018} identified attacks affecting the device \textit{firmware} throw a configuration link (e.g., USB ports), having an impact on data integrity and confidentiality, also generating disruptions on the system. Pycroft et al. \cite{pycroft:brainjacking:2016} identified the possibility of injecting malicious firmware updates. Moreover, we identify that these attacks can serve as an opportunity to generate safety problems. Ienca et al. \cite{ienca:neuroprivacy:2015, ienca:hackingBrain:2016} documented \textit{cryptographic attacks}, indicating that Cody's Emokit project was able to crack the encryption of data directly from the Emotiv EPOC, a consumer-grade \gls{BCI}. They detected that these attacks affect data integrity and confidentiality. Marin et al. \cite{Marin:securityNeurostimulators:2018} detected that current \glspl{IMD} lack robust security mechanisms. Yaqoob et al. \cite{yaqoob:cybersecurityIMD:2019} identified that neurostimulation devices lack encryption and usually define default passwords, impacting integrity and confidentially, easing unauthorized access to sensitive data. We also identify that they produce service availability and safety issues if they can modify the data. 

The authors of \cite{Camara:IMDsecurity:2015, pycroft:securityIMD:2018} highlighted that attackers could focus on draining the battery of the device and thus affect both service availability and users' physical safety. In neurostimulation systems, losing the battery capacity would result in a loss of treatment, where the disease symptoms would reappear. Due to this, some IMDs include rechargeable batteries, reducing the risks of depleting them, and thus defining more robust solutions. It is also essential to consider that, in non-rechargeable systems, surgery is required to replace the batteries, increasing the risk of both physical and psychological safety issues.

The authors of \cite{pycroft:brainjacking:2016, bonaci:appStores:2015} described the possibility of \textit{hijacking attacks}, referred to as \textit{brainjacking}, where the attacker acquires complete access over the device by any means. These attacks generate an impact on all four security impact metrics. Finally, Pycroft et al. \cite{pycroft:securityIMD:2018} identified general confidentiality impacts than can be shared by multiple attacks. They identified that close-loop \glspl{IMD} use physiological data acquired by the \gls{BCI} to improve the stimulation procedures or drug delivery. However, this sensitive data can be used by attackers to acquire information about the patient's health condition. Furthermore, an attacker can acquire sensitive information stored in the device, such as stimulation settings, personal data, or battery status, useful to perform new attacks.

Considering \glspl{NCD}, Ballarin et al. \cite{ballarin:cybersecurityAnalysis:2018} identified \textit{social engineering and phishing attacks} against \glspl{BCI}, focused on the acquisition of users' authentication credentials, affecting data confidentiality. Although \gls{BCI} applications do not require a connection to the Internet, the \gls{NCD} can be connected. Therefore, we detect that these systems can suffer \textit{malware attacks} and, specifically, \textit{ransomware} \cite{al-rimy:ransomware:2018} and those based on \textit{botnets} \cite{Kurose:computerNetworking:2017, kolias:iotBotnets:2017, stallings:cryptographySecurity:2017}, with an impact on the integrity and availability of data and applications contained in the \gls{NCD}, as well as users' safety. In particular, \textit{botnets} also generate data confidentiality issues, since attackers have control over the system. Moreover, we detect \textit{sniffing attacks} on \glspl{NCD} taking advantage of networking configuration and protocols, such as MAC flooding, DHCP attacks, ARP spoofing, or DNS poisoning \cite{anu:sniffingAttacks:2017}, affecting service and data integrity, confidentiality and availability. 

Focusing on the communication between \gls{BCI} devices and \glspl{NCD}, Sundararajan et al. \cite{Sundararajan:privacySecurityIssues:2017} studied the security of the commercial-grade Emotiv Insight, which implemented \gls{BLE} in its version 4.0 to communicate with a smartphone that contains the application offered by Emotiv. They successfully performed \textit{man-in-the-middle attacks} over the \gls{BLE} link, being able to intercept and modify information, force the \gls{BCI} to perform unwanted tasks, and conduct \textit{replay attacks} affecting, therefore, integrity, confidentiality, and availability of sensitive data. The literature has documented further integrity and confidentiality impacts, where attackers can intercept and modify sensitive data even using encryption \cite{li:bciApplications:2015, Sundararajan:privacySecurityIssues:2017, ballarin:cybersecurityAnalysis:2018, Takabi:firewallBrain:2016} \cite{pycroft:securityIMD:2018, Landau:security_EEG:2020, Marin:securityNeurostimulators:2018}. These attacks are related to the \textit{cryptographic attacks} described above, where weak encryption of the data stored in the device can derive in \textit{man-in-the-middle attacks}. Finally, it is important to note that the attacks related to user data and credentials have a higher impact if multiple users use the system. 

\subsubsection{Countermeasures}
To some of the previous attacks, different countermeasures have been proposed. Related to \textit{firmware attacks}, Ballarin et al. \cite{ballarin:cybersecurityAnalysis:2018} indicated the encryption of the firmware, as well as an authenticity verification throw hash or signature. Pycroft et al. \cite{pycroft:brainjacking:2016} highlighted periodic firmware updates and the use of authorization mechanisms for these updates. The authors of \cite{Camara:IMDsecurity:2015, pycroft:brainjacking:2016, pycroft:securityIMD:2018} identified the use of access control mechanisms placed in external devices with proximity to the patient and anomaly detection systems over the \gls{BCI} device usage to face potential threats such as \textit{battery drain attacks}. In particular, for these attacks, rechargeable batteries are recommended to avoid a surgical replacement. The authors of \cite{Landau:security_EEG:2020} proposed, as general countermeasures, the regulation of neurotechnology as a way to standardize its manufacturing processes, as well as a reduction of \gls{BCI} training process, which tends to frustrate the users, being less willing to cooperate. These measures are complementary with those documented by \cite{pycroft:securityIMD:2018}, which considered that \gls{BCI} devices should keep logs and access events, including mechanisms for reporting bugs.

The use of robust cryptographic mechanisms and the latest protocol versions are determinant to avoid \textit{cryptographic attacks}, \textit{man-in-the-middle attacks}, and \textit{sniffing attacks} \cite{Sundararajan:privacySecurityIssues:2017, ballarin:cybersecurityAnalysis:2018}. Besides, anonymization of the information transmitted from \gls{BCI} to \gls{NCD} is also recommendable against attacks impacting confidentiality, for example, using the \gls{BCI} Anonymizer \cite{Takabi:firewallBrain:2016, bonaci:appStores:2015, bonaci:exocortex:2015}. \textit{Social engineering and phishing attacks} focused on credential theft can be reduced by implementing a second authentication factor to access the \gls{BCI} and proper access control mechanisms \cite{takabi:privacyThreatsCounter:2016, ballarin:cybersecurityAnalysis:2018} \cite{pycroft:securityIMD:2018}. The application of the \textit{malware} countermeasures indicated in Section \ref{subsec:cycleProcessingConversion} can evade global \textit{malware} threats impacting \glspl{NCD}, by updating all software to the latest version and implementing periodic backup plans. Moreover, the use of \gls{ML} techniques, as proposed by Fern\'andez-Maim\'o et al. \cite{fernandezMaimo:ransomwareMCPS:2019} for \gls{MCPS}, can also be used to detect, classify, and mitigate \textit{ransomware attacks}. Concerning \textit{botnets}, a wide variety of detection techniques have been detected by us for the \gls{BCI} field, like the use of anomaly detection based on \gls{ML} and signatures, the quarantine of infected devices, and the interruption of particular communication flows \cite{amini:botnet:2015, kirubavathi:botnetAndroid:2018, mahmoud:botnet:2015}. Finally, we consider that the recommendations of the \gls{FDA} for premarket and postmarket management of security in medical devices apply to \gls{BCI} \cite{FDA:premarket:2018, Schwartz:FDAmedicalCyber:2018, FDA:postmarket:2016}. 

\subsection{Global \gls{BCI}}
\label{subsec:archGlobalBCI}

% Description of the architecture
\subsubsection{Architecture description}
Global \gls{BCI} architectures focus on the management of neural data acquisition and neural stimulation of multiple users through an Internet connection. This architecture considers three devices to deploy the phases making up the \gls{BCI} cycle, as can be seen in \figurename~\ref{fig:archBCI}. In this family, the \gls{BCI} device remains focused on data acquisition and stimulation (phase 2), whereas the \gls{NCD} is in charge of the execution of applications (phase 5), as well as conversion and processing actions (phase 3). Finally, the new element introduced in this architecture is the \gls{RCD}, representing one or more external resources or services accessible via the Internet, such as cloud computing and storage. It typically implements phases 4 and 5 of the \gls{BCI} cycle, as it has the resources to run more complex applications and information analysis. The main difference between this architecture and the one described for Local \glspl{BCI} in Section \ref{subsec:archLocalBCI} is that, in Local \glspl{BCI}, the \gls{NCD} does not send user information to external services (e.g., cloud). Finally, this section focuses on the problems associated with the communication between \gls{NCD} and \gls{RCD}, and the BCI-related attacks that can apply to \glspl{RCD}. However, these later attacks are addressed in a general way, as specific cloud computing attacks are outside the scope of this article. 

% Examples of deployments
\subsubsection{Examples of deployments}
This architectural deployment is the most innovative, as it allows the communication of multiples users with external services and the creation of complex deployments, where the data and information of every user are stored and managed in a shared infrastructure. From a commercial point of view, Emotiv allows users to contrast their data with the data stored by other users, as well as keep users' neural recordings in the cloud to visualize and manipulate them, also offering an API called Emotiv Cortex \cite{emotivCortex}. Besides, several companies worldwide provide distributed \gls{BCI} services, as is the case of Lifelines Neuro \cite{lifelinesNeuro}, which offers a continuous \gls{EEG} acquisition, storage, and visualization in their cloud platform. These scenarios are especially relevant in the context of personalized medicine and early diagnosis.

% Attacks and countermeasures
\subsubsection{Attacks and countermeasures}
Considering the attacks on this deployment, the issues documented in Section \ref{subsec:archLocalBCI} for Local \glspl{BCI} are also applicable in this architecture. However, Global \glspl{BCI} present higher risks, since these deployments are an opportunity for remote attacks against interconnected \gls{BCI} devices, which derives in physical harm for their users. Furthermore, Takabi et al. \cite{takabi:privacyThreatsCounter:2016} detected that \gls{BCI} applications could send raw brain signals to cloud services that execute \gls{ML} techniques to extract sensitive information and therefore affect confidentiality. We identify that this problem can also be present in Local \glspl{BCI} if the \gls{NCD} has an Internet connection. Ballarin et al. \cite{ballarin:cybersecurityAnalysis:2018} identified that \textit{man-in-the-middle attacks} could occur in the communication channel between \gls{NCD} and \gls{RCD}, affecting the integrity and confidentiality of the data transmitted as well as the service availability. They also detected that attacks on \glspl{RCD} could have a higher impact on confidentiality than on Local \glspl{BCI}, as these platforms store sensitive information from multiple users, that can be stolen or sold to third parties. Ienca et al. \cite{ienca:bciConsumerDevices:2018} detected different issues in Global \glspl{BCI} in terms of their usage. First, they highlighted that current brands, such as Emotiv \cite{emotiv}, indicate in their privacy policy that they can gather personal data, usage information, and interactions with other applications, and that they can infer information from these sources, with potential confidentiality issues. The authors identified as possible the use of big data to extract associations and share the data with third parties. Moreover, they detected that the use of cloud services could derive in a massive database theft with sensitive data, an unclear legal liability in case of breaches.

We identify that this architecture is quite similar to those defined and implemented for \gls{IoT} scenarios, where constrained devices communicate with external services via intermediate systems, especially when multiple devices interact. We detect that most of the security attacks and impacts defined by Stellios et al. \cite{stellios:IoTsecurity:2018} are also applicable in this architecture. Moreover, we consider that the issues highlighted by the OWASP in their \gls{IoT} projects are critical aspects of Global \glspl{BCI} \cite{OWASP:IoT:2018}. This relationship between \gls{IoT} and external services has been previously studied in cloud computing scenarios \cite{botta:cloudComputingIoT:2016}. Despite the advantages, attacks on cloud computing can impact integrity, confidentiality and availability in different cloud architecture levels, such as infrastructure, networking, storage, and software \cite{basu:cloudComputing:2018, singh:cloudComputing:2016}. The evolution of \glspl{NCD} derives in mobile devices with higher computing capabilities, integrated into mobile cloud computing systems. However, they also have an impact on the security of deployments \cite{mollah:cloudComputing:2017}. We also detect that the improvement of \glspl{NCD} capabilities can also allow the introduction of fog computing in Global \glspl{BCI}, where \glspl{NCD} perform part of the computation, generating new security and trust issues \cite{zhang:fogComputing:2018, roman:edgeComputing:2018, mahmud:fogComputing:2018}. \textit{Malware attacks} are also present in cloud environments, where ransomware and botnets are common threats \cite{singh:cloudComputing:2016}. 

Focusing on general cloud computing countermeasures, Amara et al. \cite{amara:cloudComputing:2017} identified security threats and attacks, as well as the mitigation techniques against them. The use of honeypots, firewalls, and \gls{IDS} in cloud scenarios is convenient to reduce the impact of \textit{malware attacks} \cite{roman:edgeComputing:2018}.

\begin{figure*}[htbp]
	\centering
	\includegraphics[width=\columnwidth]{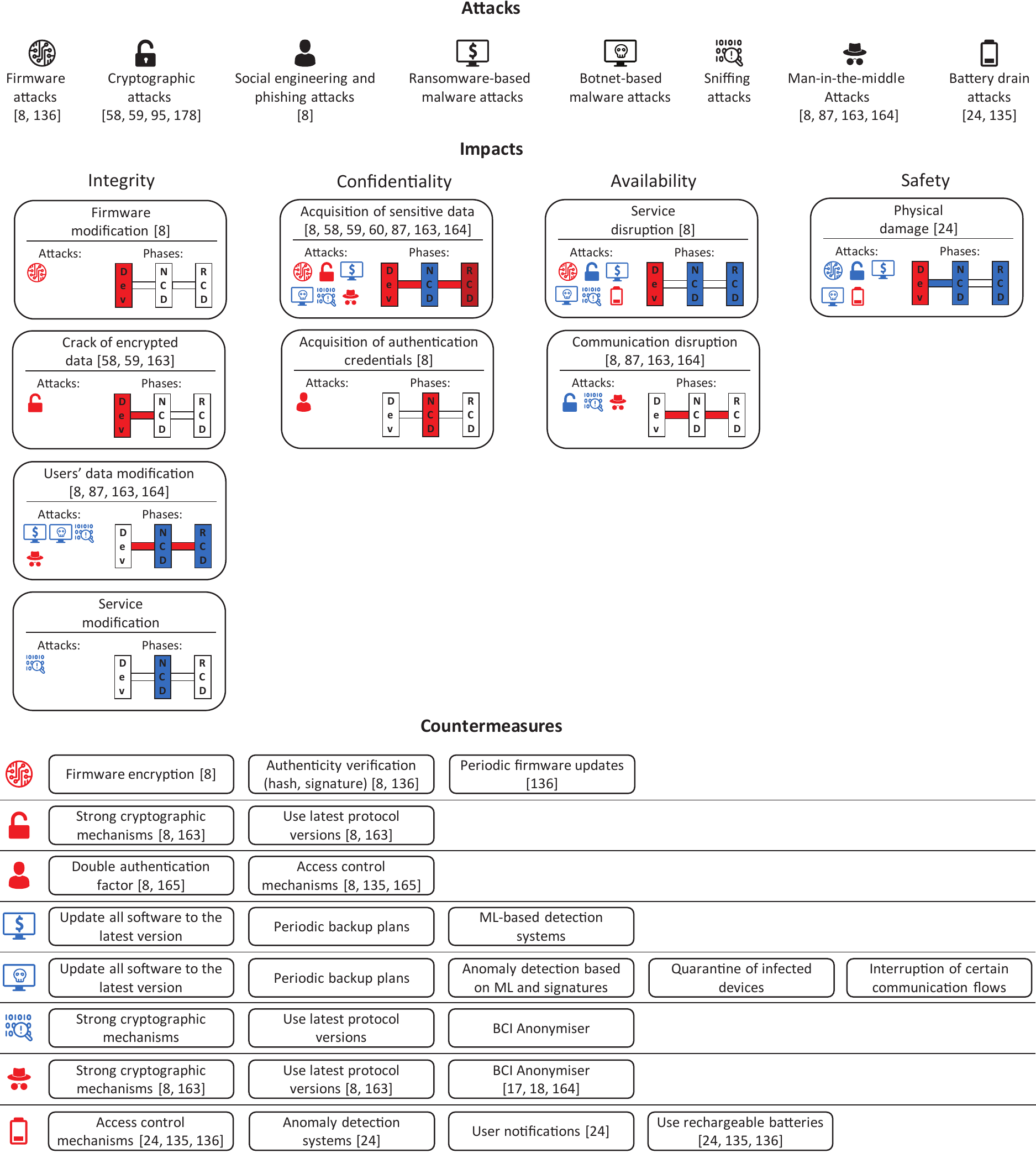}
    \caption{Attacks, impacts, and countermeasures associated with the \gls{BCI} architectural deployments. Elements indicated in red represent information detected in the literature, while blue represents our contribution.}
    \label{fig:attacksArch}
\end{figure*}

\figurename~\ref{fig:attacksArch} summarizes the previous attacks, impacts, and countermeasures. This figure first shows the list of attacks considered in this section, associated with a unique icon, where those attacks with references indicate that they have been detected in the literature, while those without references represent our contribution. After that, we show the impacts that generate the previous attacks, organized by category. For each impact, we indicate the specific attacks that cause the impact, and which elements of the architectural deployments presented in \figurename~\ref{fig:archBCI} are affected. Moreover, we consider the issues on the communication links between these elements. In particular, the attacks and elements identified in red represent issues detected in the literature, whereas those in blue are our contributions. Finally, this figure lists countermeasures detected both in the literature and by us, associating each attack with a list of countermeasures. The color and reference criteria used before for the impacts also applies to the countermeasures, where an attack represented with a particular color indicates that all their countermeasures have the same color.

\section{BCI trends and challenges}
\label{sec:bciTrendChallenges}

% Short historical review
One of the first \gls{BCI} solutions was developed at the end of the 1990s. It supposed a significant advancement in the medical industry, specifically in neurorehabilitation, bringing to the reality the mental control of prosthetic limbs and wheelchairs \cite{nicolelis:actionsThoughts:2001}. During the decade of the 2000s, a new generation of neuroprosthetic devices was developed to restore the mobility of patients severely paralyzed, creating communication links between the brain and a wide variety of actuators, such as robotic exoskeletons \cite{Lebedev:BrainMachineIF:2017}. This trend in the field of \gls{BCI} has resulted in new paradigms and scenarios in the last decade, where acquisition and stimulation procedures are used together to acquire brain activity and deliver feedback to the brain or peripheral nerves, defining the concept of bidirectional, or closed-loop, \glspl{BCI}. Focusing on these systems, NeuroPace RNS is the only technology clinically approved for closed-loop treatment \cite{Edwards:neurostimulation:2017}. DBS is nowadays considered as a unidirectional BCI system, or open-loop, only performing stimulation actions. Nevertheless, current research aims to develop closed-loop DBS systems that are able to automatically identify the best stimulation parameters based on the status of the brain \cite{Hell:DBS_closed_loop:2019}. This evolution is also applicable for neuroprostheses, where the users can mentally control prosthesis while receiving stimulation to recover motor abilities \cite{Levi:closed_loop_neuroprostheses:2018}.

\begin{figure}[htbp]
\begin{center}
\includegraphics[width=\columnwidth]{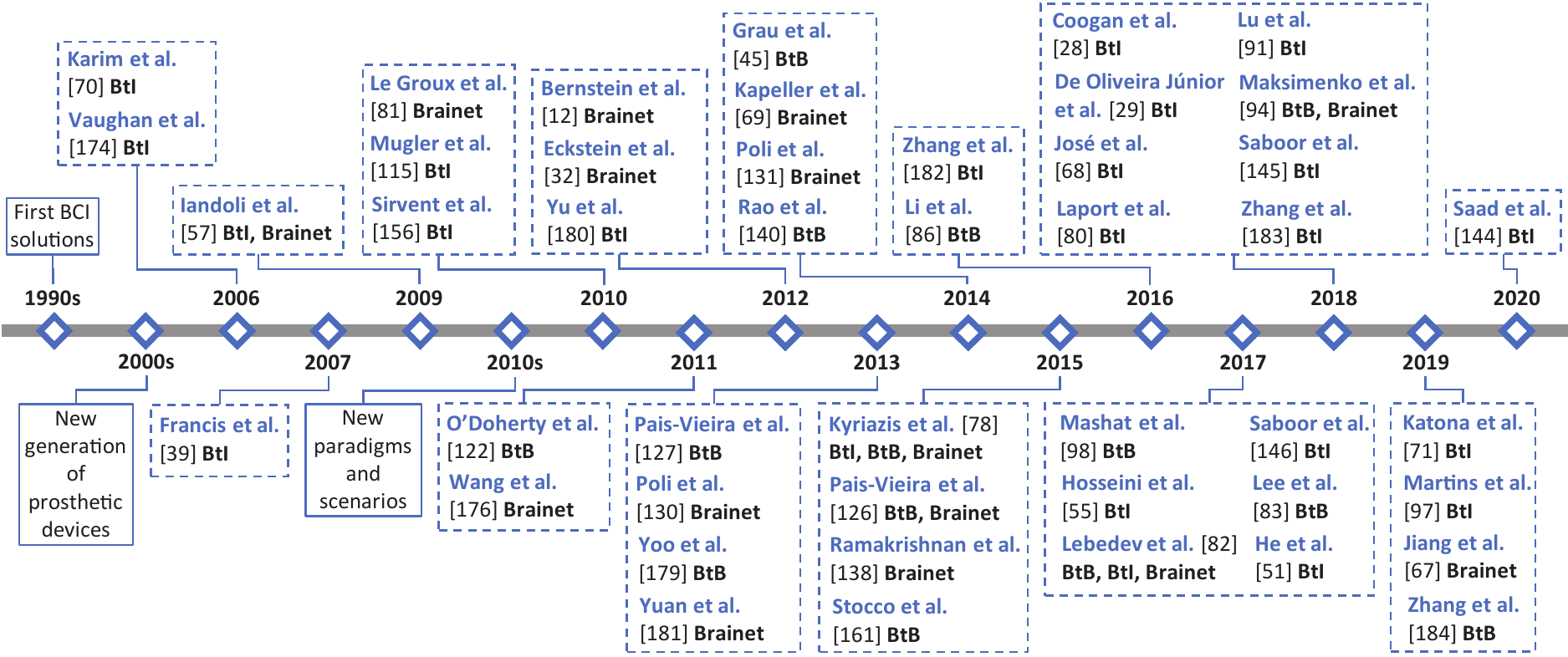}
\end{center}
\caption{Timeline of the evolution of \gls{BCI} research, seen from the perspective of \gls{BtI}, \gls{BtB}, and Brainet approaches.}
\label{fig:trends}
\end{figure}

% Prospect BCI systems
This evolution allowed the definition of prospect ways of interaction where the \gls{BCI} acts as an online communication element with other systems and users, based on Global \gls{BCI} architectures. In particular, we subsequently present several examples of futuristic systems to highlight the importance of security in the progress of \gls{BCI} technologies. Zhang et al. \cite{zhang:brainInternet:2016} defined the concept of the Internet of Brain, also known as \gls{BtI}, where the \gls{BCI} uses an \gls{NCD} to access Internet services, such as search results or social media. Lebedev et al. \cite{Lebedev:BrainMachineIF:2017} also described experiments where monkeys controlled remote robotic arms using \gls{BCI} devices. More recently, Saad et al. \cite{Saad:BtI:2020} identified that 6G technologies could enable the interconnection of \glspl{BCI} with the Internet. Besides, Martins et al. \cite{martins:bciCloud:2019} documented a fusion between neuralnanorobotics and cloud services to acquire knowledge, defining the concept of \gls{B/CI}. Another futuristic approach, \gls{BtB}, allows direct communications between two brains, known as \gls{BtB} \cite{Pais-Vieira:BtB:2013, zhang:BtB:2019}, where Pais-Vieira et al. \cite{Pais-Vieira:BtB:2013} documented the real-time exchange of information between the brain of two rats. These systems have also been extended to create networks of interconnected brains, known as Brainet, which can perform collaborative tasks between users and share knowledge, memories, or thoughts through remote brains \cite{PaisVieira:brainet:2015, jiang:brainets:2019}. Although these systems are in an early research stage, they could be a reality in the next decades, where security aspects will gain enormous importance. To represent this trend, \figurename~\ref{fig:trends} illustrates this evolution of the literature, indicating the years of publication and approaches. Besides, current innovations, such as the use of silicon-based chips, could increase the quantity of information that we can acquire from the brain, and ease the development of electronic devices to improve the resolution of the neural acquisition and sensitivity of the process \cite{Obaid:silicon-BCI:2020}.

% Projects and funding
The \gls{BCI} research field has gained relevance in the last few years, where different governments have funded and promoted \gls{BCI} initiatives. In the United States of America, the DARPA is supporting the BRAIN Initiative (Brain Research through Advancing Innovative Neurotechnologies) \cite{theBRAINinitiative}. Canada has launched its research line, called the Canadian Brain Research Strategy \cite{illes:CanadaBrain:2019, CanadaBrain}. On the other side of the Atlantic ocean, the European Union has also supported different projects, such as the Human Brain Project (HBP) \cite{humanBrainProject} or the Brain/Neural Computer Interaction (BNCI) project \cite{BNCIproject, brunner:BNCIproject:2015}. Asia has also promoted several initiatives, such as the China Brain Project \cite{poo:ChinaBrain:2016} or the Brain/MINDS project in Japan \cite{BrainJapan}. All the previous initiatives and projects aim to advance the understanding of the human brain through the use of innovative technologies. As a consequence, emerging technologies offer precise acquisition and stimulation capabilities that enable new \gls{BCI} application scenarios. The common interest in the study of the human brain and, in particular, on BCI leads to new opportunities for manufacturers, who can increase their competitiveness producing revolutionary \gls{BCI} services based on growing paradigms such as the IoT, cloud computing, and big data. This development derives in the improvement of the usability, accuracy and safety of the products, together with their expansion to non-medical economic sectors such as entertainment. The result of the above is a trend of \gls{BCI} towards Global \gls{BCI} architecture deployments, where multiple \gls{BCI} devices can communicate between them to perform collaborative tasks, based on the approaches of \gls{BtI}, \gls{BtB}, and Brainet. Once summarized the evolution of \gls{BCI} and its trend, below, we highlight the most relevant current and future challenges concerning security on \gls{BCI}.

\

\subsection{Interoperability between \gls{BCI} deployments}
Existing \gls{BCI} deployments consider isolated devices without standards to provide interoperability in terms of communication and data representation. This is the case of commercial \gls{BCI} brands and devices, which have been designed to resolve particular problems and are not compatible between them \cite{Ramadan:controlSignalsReview:2017}. Moreover, deployments integrating the communication between several \glspl{BCI} are ad hoc; that is, manufacturers design and implement them, considering only the requirements of a particular scenario. In this context, the current trend of \glspl{BCI} towards paradigms such as the \gls{IoT} and cloud computing will require an improvement in interoperability, as it is essential to ensure the future expansion of \gls{BCI} technologies. Besides, the lack of interoperability limits the definition of global cybersecurity systems and mechanisms that can be applied. In this sense, current \gls{BCI} solutions are device-oriented and do not offer collaborative mechanisms against cyberattacks. We detect as a future opportunity the use of well-known standardized APIs, communication technologies, and protocols to offer seamless protection on \gls{BCI}. We also propose the use of ontologies to represent neural information in a formal and standardized fashion. Different companies and products would use a joint representation to ease data interpretation, processing, and sharing. This homogenization would have a positive impact on cybersecurity, enabling the design and deployment of new protocols and mechanisms for the secure exchange of particular pieces of sensitive data between independent \gls{BCI} solutions. In particular, the exchange of medical information between different organizations can be accomplished using well-known standards, as is the case of the HL7 standard \cite{HL7}.

\subsection{Extensibility of \gls{BCI} designs}
Extensibility refers to the ability of \glspl{BCI} to add new functionality and application scenarios dynamically. Nowadays, \gls{BCI} devices suffer a lack of extensibility, as companies manufacture them to provide particular services on fixed application scenarios. The neural data processing is performed in a fixed way and according to predefined premises. It means that each layer making up \gls{BCI} architectures performs particular processing tasks, which can not be changed or even modified on demand \cite{Sundararajan:privacySecurityIssues:2017}. Since each application scenario has its requirements and restrictions, the trend towards Global \gls{BCI} will need new automatic and flexible architectures and processing mechanisms over the acquired neural data. These aspects also affect the security solutions that can be applied, since current constraints of \gls{BCI} systems prevent the use of reactive and adaptive defensive mechanisms to face the threats described in previous sections. In conjunction with a lack of interoperability, the security responsibilities of each phase of the architecture are predefined and cannot be extended within that element, or delegated to be performed in other systems. As a future line of work, we highlight the design of \gls{BCI} deployments that allow the implementation of most of the operations performed in software, instead of hardware, allowing developers to change the system's behavior. Another possible solution is a modular design of \gls{BCI}, including supplementary modules, according to the requirements. However, these modifications introduce new security challenges, since software developments are more prone to errors and attacks, and new modular systems will address specific challenges, such as the verification of their authenticity.

\subsection{Data protection}
Current \gls{BCI} architectures and deployments do not consider the protection of neural data and personal information, as detected in the literature \cite{sempreboni:brainInternet:2018, Ramadan:controlSignalsReview:2017, Takabi:firewallBrain:2016}. The evolution of \glspl{BCI} towards distributed scenarios with heterogeneous and ubiquitous characteristics, such as \gls{BtB} approaches, will require the storage and management of multiple users' personal and sensitive data. Because of that, future deployments should ensure that this critical information is transmitted and processed securely. Specifically, robust cryptography mechanisms need to be applied over data communication and storage, while techniques such as differential privacy or homomorphic encryption would help to ensure the anonymization of the data. Moreover, users do not have control over their privacy preferences to define who has access to the information and in which particular circumstances. Because of that, there are no specific privacy regulations to ensure that applications and external services can access only to the neural information accepted by users, nor any limitation on manufacturers or third-parties to prevent the processing of sensitive neural data without users authorization. To improve this situation, we propose policy-based solutions that allow users to define their privacy preferences based on their particular context. Besides, we propose the use of user-friendly systems that also help users proposing privacy-preserving recommendations. These initiatives must also align with the data protection law applicable in each country.

\subsection{Physical and architectural \gls{BCI} threats}
Nowadays, a considerable amount of BCI designs and deployments do not consider cybersecurity issues such as the protection of communications, processing, storage, and applications. Although some solutions include security mechanisms, like Medtronic DBS products, some aspects must be improved. In particular, these devices use proprietary telemetry protocols \cite{Medtronic:security}, which recently has led to vulnerabilities  \cite{telemetryMedtronic}. Nevertheless, companies such as Medtronic or Boston Scientific publish security bulletins when a security vulnerability affecting their devices is detected \cite{Medtronic:bulletins, BostonScientific:bulletins}, highlighting the interest that companies have on security. Moreover, the lack of \gls{BCI} standards and, specifically, cybersecurity standards, prevent the homogenization of the security solutions implemented \cite{bonaci:appStores:2015, takabi:privacyThreatsCounter:2016, Sundararajan:privacySecurityIssues:2017, Ramadan:controlSignalsReview:2017}. The expansion of \gls{BCI} will require robust dynamic cybersecurity mechanisms to face future challenges. Moreover, the development of more precise \gls{BCI} devices and the integration of a large number of devices and systems, would result in a massive production of sensitive data. In our opinion, this context could benefit the increase of vulnerable systems and communication links. To address these challenges, manufacturers should evaluate alternatives for the mitigation of cyberattacks from multiple perspectives, aiming to implement seamless cybersecurity solutions. Based on that, we propose using 5G network technologies, since they have been designed to support a significant number of devices, which are necessary for \gls{BtB} and Brainet scenarios. In particular, we identify that techniques and paradigms associated with 5G, such as \gls{NFV} and \gls{SDN} for the virtualization and dynamic management of network communications, are useful for the development of reactive cybersecurity solutions. Also, technologies such as Blockchain can provide the tracking of the information and ensure that it has not been modified, guaranteeing the integrity of the data. Moreover, we identify the protection of network communications by using protocols such as TLS \cite{IETF:TLS:2018} or IPsec \cite{IETF:IPsec:2011} as an opportunity, which offers robust mechanisms against cyberattacks. Moreover, we detect that the application of information risk management standards, such as the ISO 27000 \cite{ISO:27001}, and the NIST Cybersecurity Framework \cite{NIST:cybersecurityFramework} could benefit the creation of homogeneous and robust solutions. Finally, we identify that game theory applied to \gls{BCI} security strategies can be useful to implement regularly evolving systems. In particular, they can be useful to model how to establish the most appropriate countermeasures against continuously and automatically changing attacks, specifically in distributed scenarios such as \gls{BtB} \cite{attiah:game_theory:2018}.

\section{Conclusion}
\label{sec:conclusions}

This article performs a global and comprehensive analysis of the literature of \glspl{BCI} in terms of security and safety. Mainly, we have evaluated the attacks, impacts and countermeasures that \gls{BCI} solutions suffer from the software's architectural design and implementation perspectives. Initially, we proposed a unified version of the \gls{BCI} cycle to include neural data acquisition and stimulation processes. Once having a homogeneous \gls{BCI} cycle design, we identified security attacks, impacts, and countermeasures affecting each phase of the cycle. It served as a starting point to determine which processes and functioning stages of \glspl{BCI} are more prone to attacks. The architectural deployments of current \gls{BCI} solutions have also been analyzed to highlight the security attacks and countermeasures related to each approach to understanding the issues of these technologies in terms of network communications. Finally, we provide our vision regarding \gls{BCI} trends and depict that the current evolution of \glspl{BCI} towards interconnected devices is generating tremendous security concerns and challenges, which will increase in the near future.

Among the learned lessons, we highlight the following five: (1) the field of security oriented to \gls{BCI} technologies is not yet mature, generating opportunities for attackers; (2) even non-sophisticated attacks can have a significant impact on both \gls{BCI} technologies and users' safety; (3) there is a current opportunity for standardization initiatives to unify \glspl{BCI} in terms of information security; (4) well-studied fields, such as \glspl{IMD} and \gls{IoT}, can define a guide to develop robust security mechanisms for \glspl{BCI}; (5) users' awareness of \gls{BCI} security issues is vital.
 
As future work, we plan to focus our efforts on the design and implementation of solutions able to detect and mitigate attacks affecting the stimulation process in real time. In this context, we are considering using artificial intelligence techniques to detect anomalies in the firing patterns and neural activity controlled by \gls{BCI} solutions in charge of stimulating the brain. Besides, we also plan to contribute by improving the interoperability and data protection mechanisms of existing \gls{BCI} architectures. Finally, another future work is the development of dynamic and proactive systems as an opportunity to mitigate the impacts of the attacks documented in this work.

\section{Acknowledgments}
This work has been supported by the Irish Research Council under the government of Ireland post-doc fellowship (grant GOIPD/2018/466), by the Science Foundation Ireland (SFI) under grant number 16/RC/3948 and co-funded under the European Regional Development Fund and by FutureNeuro industry partners, and by the European Union’s Horizon 2020 Research and Innovation Programme through the Marie Skłodowska-Curie grant under Agreement 839553. Finally, we would also like to thank Mattia Zago for his advice during the development of the visual support of the work.

%% The next two lines define the bibliography style to be used, and the bibliography file.
\bibliographystyle{ACM-Reference-Format}
\bibliography{main}
\nocite{*}

\end{document}